\documentclass[10pt]{article}
\usepackage{usenixlike}
\usepackage{needspace}
\input{macros}

\hasTierCtrue

\title{\Large \bf \system: Certified Semantic Discovery for
Peer-to-Peer Agent Networks over Exact-Key DHTs}

\author{
\begin{tabular}{@{}cccc@{}}
Taotao Wang\footnotemark[2] &
Chonghe Zhao\footnotemark[2] &
Shengli Zhang &
Soung Chang Liew \\
{\small Shenzhen University} &
{\small Guangzhou University} &
{\small Shenzhen University} &
{\small The Chinese University of Hong Kong}
\end{tabular}
}
\date{}

\begin{document}
\maketitle
\renewcommand{\thefootnote}{\fnsymbol{footnote}}
\footnotetext[2]{Equal contribution. Correspondence: \texttt{ttwang@szu.edu.cn}.}
\renewcommand{\thefootnote}{\arabic{footnote}}

\begin{abstract}
\begingroup\color{black}
Tasks involving external data or operational state may
require capabilities exposed through agent endpoints or service APIs.
When a service requester is not already bound to a service provider,
it must discover advertised capabilities matching its task and
interface requirements. Over exact-key distributed hash tables (DHTs),
broad retrieval transfers large candidate lists, whereas selective
retrieval may miss relevant providers or require more replication and
lookups. Open publication also lets providers inflate their exposure
unless publication bounds are enforceable. We present \system, a
certified semantic index for discovering agent-accessible capabilities
over exact-key DHTs. A two-layer semantic sketch uses coarse cells to
group nearby descriptors and residual codes to narrow candidate
selection. Service providers publish at a bounded set of derived keys,
while requesters probe precision keys before broader recall keys within
a lookup budget. Anchor committees certify each descriptor's
publication-key set, enabling storage services and requesters to enforce
descriptor-to-key consistency. On real API descriptors and task queries,
\system achieves $\mathrm{recall}@10=0.955$ against exact embedding-space
neighbors and $0.947$ against relevance labels from ToolBench, a benchmark
for language-model tool use. On a corpus with controlled density
augmentation, it matches the candidate exposure of tuned
locality-sensitive hashing (LSH) over a DHT at recall $0.95$ with
$7.7\times$ fewer lookups and reduces publication fan-out from 16 to 10.
A Go/libp2p prototype deployed on same-region and cross-region 200-peer
cloud overlays replays 299 Internet queries. With parallel probes and
cold certificate caches, \system achieves mean completion-time speedups
of $3.64\times$ and $4.11\times$ over LSH, respectively.
\endgroup
\end{abstract}

\noindent\parbox{\linewidth}{\textbf{Keywords:}
agent-accessible services; semantic capability discovery;
distributed hash tables; certified publication.}

\section{Introduction}
\label{sec:intro}
\begingroup\color{black}

An agent carrying out a task may need data or operational
state held by an external provider~\cite{mcp2025tools}. Consider a user arriving in an
unfamiliar city who asks their personal assistant agent, ``Find an
eye clinic with an appointment available tomorrow.'' The agent can
interpret the request using a language model, but must obtain current
appointment availability from a clinic's service. The clinic may expose
its appointment-query capability through an appointment assistant agent
or a structured service API~\cite{a2a2025spec,mcp2025tools}.
In either case, the user's personal
assistant acts as the service requester and needs access to an
externally provided capability.

Preconfigured services suffice when the required provider
and interface are known. We study requests for which suitable providers
must first be identified from independently published capability
descriptions~\cite{guo2026agentdiscovery}. In the appointment example,
the personal assistant first discovers services advertising the relevant query
capability, then contacts candidates to check location and current
availability. This example illustrates discovery before service interaction;
the index returns candidate descriptions, while subsequent interaction
uses the selected service's interface and authorization requirements.

We consider this discovery problem in peer-to-peer (P2P)
agent networks with independently managed providers
~\cite{wang2025ioafundamentals,wang2026agenticp2p}.
Centralized search
makes discovery depend on one service, while flooding incurs work
that grows with the network. \nsclarify{We seek a decentralized semantic
index that discovers capabilities from task intent expressed in natural
language, while bounding publication fan-out and budgeting network lookups.}
Distributed hash tables (DHTs) provide exact-key routing to responsible
peers without a central directory
~\cite{stoica2003chord,maymounkov2002kademlia,rhea2005opendht}.
\nsclarify{Recent agent-discovery systems match task and capability
semantics~\cite{guo2026agentdiscovery,xu2026agentosi,zhang2026distributedagents}.
Such matching depends on semantic similarity, whereas a DHT lookup
requires an exact key. DHT routing alone does not identify
providers whose capabilities match a task.}

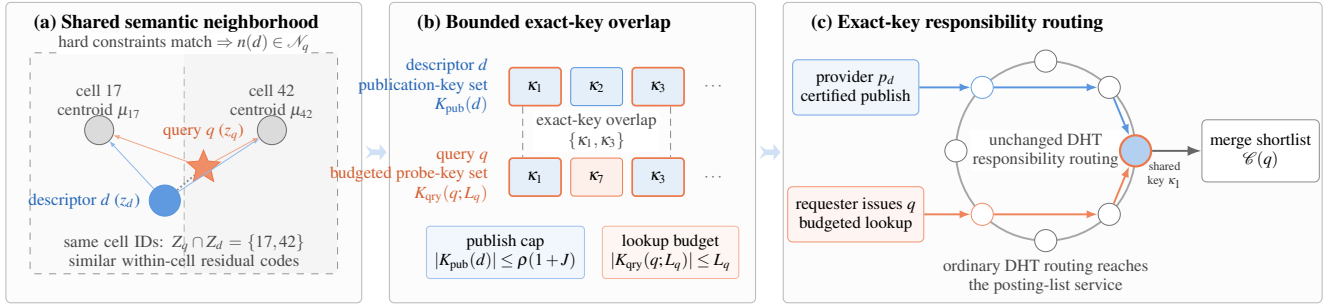
\begin{figure*}[t]
\centering
\resizebox{0.98\textwidth}{!}{%
\begin{tikzpicture}[
  panel/.style={draw=black!28,rounded corners=2pt,fill=black!1},
  ptitle/.style={font=\bfseries\footnotesize,anchor=west},
  note/.style={font=\scriptsize,text=black!72,align=center},
  chip/.style={draw=black!35,rounded corners=1.5pt,minimum width=7.5mm,
               minimum height=5.5mm,font=\scriptsize,fill=white},
  providerchip/.style={chip,draw=vizblue!80,fill=vizblue!12},
  querychip/.style={chip,draw=vizorange!85,fill=vizorange!10},
  sharedchip/.style={chip,draw=vizorange!90,line width=0.7pt,
                     fill=vizblue!12},
  peer/.style={circle,draw=black!55,fill=white,minimum size=3.2mm,
               inner sep=0pt},
  flow/.style={-{Latex[length=2.0mm]},thick,black!60},
  bridge/.style={draw=none,fill=vizblue!22}
]

\draw[panel] (0,0) rectangle (5.15,4.35);
\node[ptitle] at (0.28,4.05) {(a) Shared semantic neighborhood};

\fill[black!2] (0.36,0.43) rectangle (2.57,3.61);
\fill[black!4] (2.58,0.43) rectangle (4.79,3.61);
\draw[black!42,dashed,rounded corners=2pt] (0.35,0.42) rectangle (4.80,3.62);
\draw[black!30,densely dashed] (2.58,0.50) -- (2.58,3.48);
\node[font=\scriptsize,fill=white,inner sep=0.5pt,text=black!72]
  at (2.58,3.76) {hard constraints match $\Rightarrow n(d)\in\mathcal N_q$};

\node[circle,draw=black!58,fill=black!14,minimum size=4.2mm,
      inner sep=0pt] (mu17) at (1.35,2.50) {};
\node[font=\scriptsize,text=black!68,above=1pt,align=center] at (mu17)
  {cell 17\\centroid $\mu_{17}$};
\node[circle,draw=black!58,fill=black!14,minimum size=4.2mm,
      inner sep=0pt] (mu42) at (3.85,2.50) {};
\node[font=\scriptsize,text=black!68,above=1pt,align=center] at (mu42)
  {cell 42\\centroid $\mu_{42}$};

\node[circle,draw=vizblue!80,fill=vizblue!75,minimum size=4.5mm,
      inner sep=0pt,label={[font=\scriptsize,text=vizblue!85!black]left:descriptor $d$ ($z_d$)}]
  (desc) at (2.30,1.48) {};
\node[star,star points=5,star point ratio=2.2,draw=vizorange!90,
      fill=vizorange!75,minimum size=5.0mm,inner sep=0pt,
      label={[font=\scriptsize,text=vizorange!90!black]above:query $q$ ($z_q$)}]
  (query) at (2.86,1.96) {};
\draw[black!45,densely dotted,thick] (desc) -- (query);
\draw[-{Latex[length=0.9mm]},vizblue!52] (desc) -- (mu17);
\draw[-{Latex[length=0.9mm]},vizblue!52] (desc) -- (mu42);
\draw[-{Latex[length=0.9mm]},vizorange!58] (query) -- (mu17);
\draw[-{Latex[length=0.9mm]},vizorange!58] (query) -- (mu42);
\node[note,text width=43mm] at (2.58,0.76)
  {same cell IDs: $Z_q\cap Z_d=\{17,42\}$\\
   similar within-cell residual codes};

\begin{scope}[xshift=-0.20cm]
\draw[panel] (5.75,0) rectangle (11.05,4.35);
\node[ptitle] at (6.03,4.05) {(b) Bounded exact-key overlap};

\node[font=\scriptsize,align=right,anchor=east,text=vizblue!85!black] at (7.30,3.13)
  {descriptor $d$\\publication-key set\\$K_{\mathrm{pub}}(d)$};
\node[sharedchip]   (p1) at (7.86,3.13) {$\kappa_1$};
\node[providerchip] (p2) at (8.75,3.13) {$\kappa_2$};
\node[sharedchip]   (p3) at (9.64,3.13) {$\kappa_3$};
\node[font=\scriptsize,text=black!55]              at (10.45,3.13) {$\cdots$};

\node[font=\scriptsize,align=right,anchor=east,text=vizorange!90!black] at (7.30,1.82)
  {query $q$\\budgeted probe-key set\\$K_{\mathrm{qry}}(q;L_q)$};
\node[sharedchip] (q1) at (7.86,1.82) {$\kappa_1$};
\node[querychip]       at (8.75,1.82) {$\kappa_7$};
\node[sharedchip] (q3) at (9.64,1.82) {$\kappa_3$};
\node[font=\scriptsize,text=black!55]              at (10.45,1.82) {$\cdots$};

  \draw[black!45,densely dashed]
    ($(p1.south)+(-0.18,0)$) -- ($(q1.north)+(-0.18,0)$);
  \draw[black!45,densely dashed]
    ($(p3.south)+(0.18,0)$) -- ($(q3.north)+(0.18,0)$);
\node[font=\scriptsize,text=black!72,fill=white,inner sep=1pt,align=center]
  at (8.75,2.48) {exact-key overlap\\$\{\kappa_1,\kappa_3\}$};

\node[draw=vizblue!45,rounded corners=1.5pt,fill=vizblue!6,
      font=\scriptsize,align=center] at (7.43,0.73)
  {publish cap\\$|K_{\mathrm{pub}}(d)|\leq\rho(1+J)$};
\node[draw=vizorange!50,rounded corners=1.5pt,fill=vizorange!6,
      font=\scriptsize,align=center] at (9.82,0.73)
  {lookup budget\\$|K_{\mathrm{qry}}(q;L_q)|\leq L_q$};
\end{scope}

\draw[panel] (11.25,0) rectangle (19.05,4.35);
\node[ptitle] at (11.53,4.05) {(c) Exact-key responsibility routing};

\node[draw=vizblue!65,rounded corners=2pt,fill=vizblue!7,
      minimum width=15mm,minimum height=8mm,font=\scriptsize,align=center]
  (provider) at (12.30,3.119) {provider $p_d$\\certified publish};
\node[draw=vizorange!70,rounded corners=2pt,fill=vizorange!7,
      minimum width=15mm,minimum height=8mm,font=\scriptsize,align=center]
  (requester) at (12.30,1.281) {requester issues $q$\\budgeted lookup};

\coordinate (ctr) at (15.05,2.20);
\draw[black!35,thick] (ctr) circle (1.30);
\node[note,fill=white,inner sep=1.2pt] at (15.05,2.20)
  {unchanged DHT\\responsibility routing};

\node[peer] (n180) at ($(ctr)+(180:1.30)$) {};
\node[peer,draw=vizblue!70] (n135) at ($(ctr)+(135:1.30)$) {};
\node[peer] (n90) at ($(ctr)+(90:1.30)$) {};
\node[peer] (n45) at ($(ctr)+(45:1.30)$) {};
\node[peer,draw=vizorange!90,fill=vizblue!35,line width=0.9pt,
      minimum size=4.5mm] (owner1) at ($(ctr)+(0:1.30)$) {};
\node[font=\tiny,text=black!78,align=center,inner sep=0pt]
    at ($(owner1)+(0.42,-0.35)$) {shared\\key $\kappa_1$};
\node[peer] (n315) at ($(ctr)+(315:1.30)$) {};
\node[peer] (n270) at ($(ctr)+(270:1.30)$) {};
\node[peer,draw=vizorange!75] (n225) at ($(ctr)+(225:1.30)$) {};

\draw[-{Latex[length=1.6mm]},thick,vizblue!78] (provider.east) -- (n135.west);
\draw[-{Latex[length=1.6mm]},thick,vizblue!78] (n135) -- (n45);
\draw[-{Latex[length=1.6mm]},thick,vizblue!78] (n45) -- (owner1);

\draw[-{Latex[length=1.6mm]},thick,vizorange!82] (requester.east) -- (n225.west);
\draw[-{Latex[length=1.6mm]},thick,vizorange!82] (n225) -- (n315);
\draw[-{Latex[length=1.6mm]},thick,vizorange!82] (n315) -- (owner1);

\node[draw=black!45,rounded corners=2pt,fill=white,minimum width=15mm,
  minimum height=9mm,font=\scriptsize,align=center]
  (short) at (18.16,2.20) {merge shortlist\\$\cand{q}$};
\draw[flow] (owner1.east) -- (short.west);

\node[note] at (15.05,0.38)
  {ordinary DHT routing reaches\\the posting-list service};

\path[bridge]
  (5.20,2.08) -- (5.30,2.14) -- (5.41,2.14) -- (5.41,2.06) --
  (5.51,2.18) -- (5.41,2.30) -- (5.41,2.22) --
  (5.30,2.22) -- (5.20,2.28) -- (5.27,2.18) -- cycle;
\path[bridge]
  (10.90,2.08) -- (11.00,2.14) -- (11.11,2.14) -- (11.11,2.06) --
  (11.21,2.18) -- (11.11,2.30) -- (11.11,2.22) --
  (11.00,2.22) -- (10.90,2.28) -- (10.97,2.18) -- cycle;

\end{tikzpicture}%
}
\caption{\system compiles semantics before the DHT. (a) Matching hard
constraints ensure that the descriptor's namespace label $n(d)$ belongs
to the query's allowed namespace-label set $\mathcal N_q$; nearby
embeddings $z_q,z_d$ yield overlapping coarse-cell ID sets $Z_q,Z_d$ and
similar residual codes. (b) These codes become exact keys: the
publication-key set $K_{\mathrm{pub}}(d)$ and budgeted probe-key set
$K_{\mathrm{qry}}(q;L_q)$ overlap on $\kappa_1,\kappa_3$. Publication
uses at most $\rho(1+J)$ keys, where $\rho$ is the number of selected
coarse cells and $J$ is the number of residual-code families (one recall
publication key plus $J$ precision publication keys per coarse cell); queries use at most $L_q$.
(c) Publication and lookup
for each shared key follow unchanged exact-key routing to the same
responsible peer, whose posting-list service returns postings for local merge.
The ring denotes ownership only, not Chord;
evaluation uses Kademlia-style routing.}
\label{fig:teaser}
\end{figure*}

Locality-sensitive hashing (LSH) supplies a direct construction:
publish each descriptor under several LSH-derived keys and probe the
query's own and neighboring hash codes
~\cite{gionis1999lsh,lv2007multiprobe,zhu2007semanticdht,haghani2009distributedlsh,kraus2015nearbucket}.
\nsclarify{Coarse codes expose large candidate lists; finer codes can
separate related descriptors and require more publication keys or
probes to recover them. This couples retrieval quality to the amount
of replicated index state and remote lookup work. Open publication
adds an enforcement requirement: providers can increase exposure by
placing descriptors at additional high-traffic keys. Storage services
must therefore verify that each posting's target key follows from
its descriptor before accepting it.}

\nsclarify{We present \system, a certified semantic index that bounds
publication fan-out and budgets network lookups for
agent-accessible capability discovery
over exact-key DHTs.} Namespace labels encode interface and
execution constraints; requesters use task semantics to distinguish
capabilities within that compatibility scope. A two-layer semantic
sketch maps descriptor and task embeddings to exact keys. Coarse cells organize nearby
capabilities, and residual codes narrow candidate selection within
each cell. Providers publish at a bounded set of keys derived from
the sketch, while requesters probe precision keys before broader
recall keys within a physical lookup budget. Overlapping publication
and probe keys expose the provider's posting to the requester,
as illustrated in Figure~\ref{fig:teaser}.

An anchor committee independently recomputes each descriptor's
publication-key set and issues a membership certificate binding the
set to the descriptor. \nsclarify{Storage services and requesters verify
the certificate and Merkle inclusion proof. These checks
make the per-descriptor publication bound enforceable at storage
without requiring every replica to run the semantic encoder.}
\nsclarify{Ordinary DHT routing locates responsible peers,}
whose posting-list services store multiple certified postings per key.
\nsclarify{Stable lineage handles and physical sharding extend the design
to descriptor updates, revocation, and changing storage demand.}

On real API descriptors and task queries, \system achieves
$\mathrm{recall}@10=0.955$ against exact embedding-space neighbors
and $0.947$ against relevance labels from ToolBench, a benchmark for
language-model tool use~\cite{qin2023toolllm}. On a corpus with
controlled density augmentation, it matches the candidate exposure
of the best configurations in our LSH-on-DHT sweep at recall $0.95$ with
$7.7\times$ fewer exact-key lookups and reduces publication fan-out
from 16 to 10. \nsclarify{We implement a Go/libp2p prototype and deploy
two 200-peer overlays on cloud virtual machines (VMs), one within a
region and one across eight regions. We replay 299 queries over the Internet, measuring
routing, paged reads, verification, and candidate reconstruction. Both systems
use the same posting format and replica policy. With parallel probes
and cold certificate caches, \system uses 130.91 application RPCs
per query versus 777.85 for LSH, with similar message volume.
Fewer RPCs yield mean completion-time speedups
of $3.64\times$ and $4.11\times$ within and across regions.}
Separate simulations evaluate publication membership
under attack, descriptor migration, and physical sharding.

\nsclarify{This paper makes three contributions: (1) a semantic key
construction and probe schedule that connect retrieval quality to
publication fan-out and a physical lookup budget, with a precision-key
hit-probability analysis; (2) a certified publication protocol that
makes descriptor-to-key consistency verifiable at storage and retrieval;
and (3) a Go/libp2p implementation and an evaluation on cloud VMs over
the Internet, showing how index selectivity, replicated reads, and
verification affect network lookup work and completion time.}
\endgroup

\section{\nsrevision{Problem and Overview}}
\label{sec:background}
\begingroup\color{black}

\begin{figure*}[t]
\centering
\resizebox{\textwidth}{!}{%
\begin{tikzpicture}[
  every node/.style={font=\footnotesize,text=black,align=center},
  box/.style={draw=black!35,rounded corners=2pt,fill=white,
              text width=2.85cm,minimum height=1.55cm,inner sep=4pt},
  pub/.style={box,draw=vizblue!75,fill=vizblue!5},
  qry/.style={box,draw=vizorange!85,fill=vizorange!5},
  pflow/.style={-{Latex[length=1.8mm]},line width=0.85pt,draw=vizblue},
  qflow/.style={-{Latex[length=1.8mm]},line width=0.85pt,draw=vizorange},
  note/.style={font=\scriptsize,text=black!65,fill=white,inner sep=1pt}
]
\node[font=\bfseries\footnotesize,anchor=west] at (0.1,4.65)
  {Semantic keys, certified publication, and budgeted lookup};

\node[pub] (provider) at (1.65,3.25)
  {\emph{Provider}\\Clinic service\\assistant agent / API\\$n(d),x_d$};
\node[pub] (dkeys) at (5.10,3.25)
  {Semantic sketch\\coarse cells\\residual codes\\$K_{\mathrm{pub}}(d)$};
\node[pub,text width=3.0cm] (committee) at (8.80,3.25)
  {\emph{Anchor committee}\\recompute $K_{\mathrm{pub}}(d)$\\issue $\Cert_d,\Sigma_d$};

\node[qry] (requester) at (1.65,0.85)
  {\emph{Service requester}\\Personal assistant\\Eye-clinic query\\$\mathcal N_q,x_q,L_q$};
\node[qry] (qkeys) at (5.10,0.85)
  {Query sketch\\allowed namespaces\\coarse cells\\residual codes};
\node[qry,text width=3.0cm] (schedule) at (8.80,0.85)
  {Probe stages $P_1,P_2,P_3$\\precision before recall\\$K_{\mathrm{qry}}(q;L_q)$};

\draw[draw=black!40,rounded corners=2pt,fill=black!1]
  (11.10,-0.05) rectangle (14.35,4.05);
\node[font=\bfseries\footnotesize] at (12.725,3.40) {Exact-key DHT};
\node[text width=2.8cm] at (12.725,2.65)
  {route each key to\\responsible peers};
\draw[-{Latex[length=1.5mm]},draw=black!55,line width=0.7pt]
  (12.725,2.25) -- (12.725,1.65);
\node[box,text width=2.65cm,minimum height=1.55cm] (store) at (12.725,0.85)
  {\emph{Posting-list service}\\verify and store\\postings per key};

\node[qry,text width=2.8cm] (output) at (16.25,0.85)
  {Verify and merge\\fetch descriptors\\rank shortlist $\cand{q}$};
\node[text width=2.8cm,font=\scriptsize,text=black!65] at (16.25,3.18)
  {Two cells, two families:\\six publication keys\\\smallskip
   Every lookup consumes\\the query's budget $L_q$.};

\draw[pflow] (provider.east) -- (dkeys.west);
\draw[pflow] (dkeys.east) -- (committee.west);
\draw[pflow] (provider.south) -- (1.65,2.10) -- (10.65,2.10)
  -- (10.65,3.25) -- (11.10,3.25);
\draw[qflow] (requester.east) -- (qkeys.west);
\draw[qflow] (qkeys.east) -- (schedule.west);
\draw[qflow] (schedule.east) -- (11.10,0.85);
\draw[qflow] (store.east) -- (output.west);

\draw[pflow,arrows={Latex[length=1.8mm]-Latex[length=1.8mm]}]
  (provider.north) -- (1.65,4.25) -- (8.80,4.25) -- (committee.north);
\node[note,text=blue] at (5.10,4.25)
  {$\bar d$ to committee; $\Cert_d,\Sigma_d$ to provider};
\node[note,text=blue] at (5.10,2.10)
  {provider publishes certified postings};
\node[note,text=vizorange!90!black] at (8.80,-0.20) {budgeted probe keys};
\end{tikzpicture}%
}
\caption{\nsrevision{Publication and lookup share the semantic-key
construction. Namespace labels select compatible providers; coarse
cells and residual codes match tasks within that scope. The committee
recomputes publication keys from the complete descriptor and certifies
their membership. \nsclarify{The provider publishes the certified postings.}
Exact-key DHT routing reaches posting-list services;
the requester verifies returned postings and reconstructs candidates.
The clinic advertises an appointment-query capability
through an appointment assistant agent or a service API. The user's
personal assistant is the service requester; it discovers candidates
under supported interface contracts and queries live slots afterward.}}
\label{fig:pipeline}
\end{figure*}
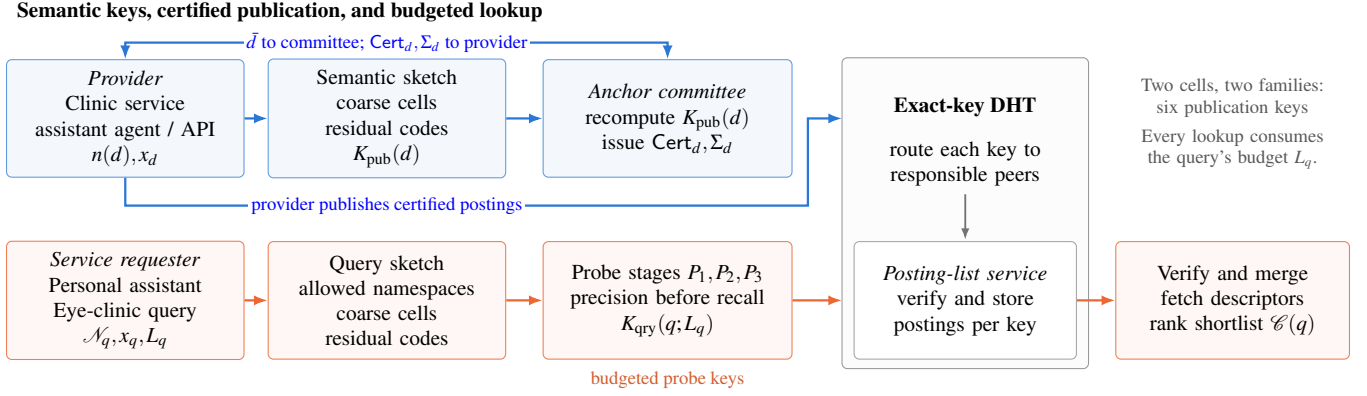

\system connects four participants: a \emph{service provider} advertises a
descriptor, an \emph{anchor committee} certifies its publication keys,
\emph{posting-list services} store postings at responsible peers, and a
\emph{service requester}
retrieves candidates. Figure~\ref{fig:pipeline}
follows their publication and query paths.

A service provider is the authenticated principal publishing a
capability description. It exposes that capability through an agent
endpoint or a service API callable by agents. In the appointment
example, the clinic is the provider and the user's personal assistant
agent is the service requester. A requester may perform discovery
through a client acting on its behalf. DHT peers supply routing and posting-list
storage; they need not themselves execute the advertised capabilities.

\subsection{Discovery and Lookup Budget}
\label{sec:background:formal}
\label{sec:background:orthogonal}
A provider advertises a descriptor version $d$ containing compatibility
conditions, semantic text $x_d$, and a retrieval URI for the complete
descriptor $\bar d$.
This URI locates the description; it is not, by
definition, the service's invocation endpoint.
The namespace label $n(d)$ encodes the compatibility
conditions to delimit the compatible search scope.
A requester issues $q=(\mathcal N_q,x_q,k_q,L_q)$,
specifying allowed namespace labels $\mathcal N_q$, task text $x_q$,
shortlist size $k_q$, and physical lookup budget $L_q$.

In the appointment example, the clinic publishes a
descriptor version $d$ whose semantic text $x_d$ describes eye-clinic
appointment queries; its complete descriptor $\bar d$ records the
interface contract and other compatibility conditions summarized by
$n(d)$. Discovery has two parts.
First, the assistant chooses $\mathcal N_q$ from the namespaces
supported by its client or adapters, such as one whose interface
accepts a department and date and returns appointment records.
A candidate must satisfy $n(d)\in\mathcal N_q$.
Second, the assistant expresses ``Find a service for querying
eye-clinic appointments'' in $x_q$; semantic matching compares this
intent with $x_d$ to find relevant services within that scope.
After discovery, the assistant checks location and other task constraints,
obtains any required authorization, and invokes a selected service
with the requested date to learn whether appointments are available
tomorrow. The indexed text $x_d$ advertises the query capability;
current slots come from the service.

An eye clinic and a dental clinic can publish
descriptors with the same namespace label $n(d)$
if they use the same interface contract, admission class, and
execution-policy class; their respective texts $x_d$ distinguish their
specialties. Conversely, an eye clinic offering a different
conversational interface requires support for that contract before it
can be included through $\mathcal N_q$. Whether the provider exposes
an agent or an API does not by itself determine namespace membership:
the compatibility conditions do.

Let $\mathcal D$ denote the descriptor population, the set of descriptor
versions indexed by the system. The \nsclarify{namespace-compatible population} for query $q$ is
\begin{equation}
\mathcal D_q=\{d\in\mathcal D:n(d)\in\mathcal N_q\}.
\label{eq:eligible-population}
\end{equation}
\system retrieves and ranks candidates from $\mathcal D_q$
using semantic matching.

To retrieve candidates, \system maps the query to exact keys
and accesses their posting lists through DHT lookups.
A DHT routes an exact key to responsible peers in
$O(\log N_{\mathrm{net}})$ hops, where $N_{\mathrm{net}}$ is the peer
count~\cite{stoica2003chord,maymounkov2002kademlia}. Those peers run a
posting-list service because many descriptors can share a key.
\nsclarify{An entry point is a peer through which the requester initiates
a DHT lookup. Each exact-key lookup consumes one unit of $L_q$.}
Looking up the same key through another entry point,
or looking up a physical shard key, consumes an additional unit.
Routing hops, replica RPCs, and page requests within a lookup do not
consume additional units of $L_q$; we measure their communication
and processing costs separately.
Thus, $L_q$ bounds the number of physical lookups,
while $k_q$ specifies the desired shortlist size. If the budgeted
lookups yield fewer than $k_q$ eligible candidates, the requester
returns a shorter shortlist.

\subsection{Publication and Query Workflow}
\label{sec:overview}
\label{sec:overview:example}
\label{sec:overview:flow}
For the clinic service provider introduced above, publication
begins by encoding its semantic text $x_d$ into an embedding.
The provider constructs a semantic sketch by selecting nearby
coarse cells and computing residual codes within each cell.
Suppose it selects two cells and uses two residual-code families.
Following the key construction described in Section~\ref{sec:index},
it then derives a publication-key set $K_{\mathrm{pub}}(d)$ containing
six keys: one recall key and two precision keys per cell.
The recall key groups descriptors assigned
to that cell, while each precision key selects those sharing a
residual code.

The anchor committee independently derives $K_{\mathrm{pub}}(d)$ from
the complete descriptor $\bar d$ and issues a membership certificate
$\Cert_d$ binding the descriptor to this publication-key set.
It authenticates $\Cert_d$ with a committee signature $\Sigma_d$.
At each key, the provider
publishes a posting containing a provider-signed posting body, the
membership certificate $\Cert_d$, the committee signature $\Sigma_d$, and a Merkle
inclusion proof that the key belongs to the certified set.
\system uses ordinary DHT routing to locate the peers responsible
for each key. Their posting-list services verify the posting
before storing it.

The user's personal assistant, acting as the service requester,
specifies the supported namespace labels $\mathcal N_q$ and expresses
its discovery task in $x_q$, such as ``Find a service for querying
eye-clinic appointments.''
It encodes $x_q$ and constructs a semantic sketch using the same
configuration as the provider. If their sketches select a common
cell and agree on a residual code, the requester derives a precision
key at which the provider has published its posting.
The requester orders its probes in three stages: exact precision keys
in $P_1$, neighboring-code and additional-cell precision keys in $P_2$,
and recall keys in $P_3$. The keys selected within the physical lookup
budget $L_q$ form the budgeted probe-key set $K_{\mathrm{qry}}(q;L_q)$.
\nsclarify{It then verifies returned postings, merges them by descriptor commitment,}
fetches complete descriptors, and ranks eligible candidates to produce
the candidate shortlist $\cand{q}$,
following the verification and ranking procedure described in
Section~\ref{sec:poison}.

\subsection{LSH-on-DHT}
\label{sec:background:baseline}
\label{sec:index:baseline}
Locality-sensitive hashing (LSH) maps similar vectors to the same
hash code with higher probability than dissimilar vectors
~\cite{gionis1999lsh}. LSH-on-DHT uses these hash codes to construct
exact keys: providers publish descriptors under keys derived from
their embeddings, and requesters probe keys derived from query
embeddings to retrieve candidates
~\cite{zhu2007semanticdht,haghani2009distributedlsh,kraus2015nearbucket}.
Longer hash codes improve selectivity; more independently seeded
hash functions or neighboring-code probes recover recall at
additional publication and lookup cost~\cite{lv2007multiprobe}.
Our evaluation tunes these choices at matched recall, using the
baseline key construction detailed in Appendix~\ref{app:lsh-keys}.
\endgroup

\section{\nsrevision{Semantic Index}}
\label{sec:index}
\begingroup\color{black}
\system's semantic index maps a descriptor to a bounded publication-key set
and a query to an ordered probe sequence. Namespace labels define the
compatibility scope; the semantic sketch determines which descriptors within
that scope are likely to share the query's keys.

\subsection{Descriptors and Compatibility}
\label{sec:index:commitment}
\label{sec:index:namespace}
Let $\bar d$ be the complete descriptor for version $d$. It records
the provider public key $\field{pk}_{p_d}$, a provider-local capability
identifier, compatibility fields, semantic text $x_d$, metadata, and a
retrieval address $\field{ptr}_d$. With cryptographic hash $H$ and
deterministic canonicalization $\canon$, its commitment is
\begin{equation}
C_d=H(\canon(\bar d)).
\label{eq:commitment}
\end{equation}
The requester verifies a fetched descriptor against $C_d$.
Whereas $C_d$ identifies one descriptor version, a stable lineage
handle $\iota_d$, derived from the provider key and capability
identifier, identifies the same capability across versions.
The anchor committee assigns a lineage epoch $e_d$ to each version
to order versions within that lineage, enabling requesters to
distinguish newer certified versions from older ones.
Appendix~\ref{app:descriptor-format} gives the complete descriptor
format for $\bar d$.

We collect the descriptor's hard compatibility requirements in a
namespace label, defined as
\begin{equation}
n(d)=(\tau_d,\sigma_d,\psi_d),
\label{eq:namespace}
\end{equation}
where $\tau_d$ is an admission class, $\sigma_d$ identifies a compatible
input/output contract, and $\psi_d$ is an execution-policy class.
For the appointment-query interface introduced in
\S\ref{sec:background:formal}, an illustrative label uses
$\tau_d=\text{\code{generic}}$, which adds no
admission distinction, $\sigma_d=\text{\code{appointment-query-v1}}$
identifies the department/date request and appointment-record response
contract, and $\psi_d=\text{\code{web-enabled}}$ permits external
network access. These values specify the compatibility scope;
the eye-clinic specialization remains in $x_d$.
\nsclarify{The versioned namespace registry $\mathcal R_\nu$ contains
canonical labels whose fields use protocol identifiers; compatible
schema variants share an interface contract, as specified in
Appendix~\ref{app:namespace-registry}.} Fields enter the label
when a mismatch prevents compatible execution or violates policy.
Task specializations and preferences remain in $x_d$; adding a
capability under an existing contract requires no new label.

A resolver maps the query's explicit hard requirements to
$\mathcal N_q\subseteq\mathcal R_\nu$, retaining the task wording in
$x_q$. Providers, anchor committees, and requesters share a signed,
content-addressed semantic-index configuration, identified by the
semantic-index configuration ID $\nu$. The configuration specifies the
namespace registry, canonical label encoding and compatibility map,
tokenizer and encoder weights,
canonical input and numeric-normalization rules, codebook centroids,
and residual-code projection seeds. \nsclarify{A deterministic reference
implementation or canonical quantization rule ensures that identical
inputs yield identical labels, cell IDs, and residual codes across
participants. Queries and descriptors are encoded separately.}

\subsection{Coarse Cells and Residual Codes}
\label{sec:index:sketch}
\label{app:reference-sketch}
To support semantic matching within the compatibility scope
defined by namespace labels, the provider converts $x_d$
into a semantic sketch. The sketch comprises coarse-cell IDs
that group nearby embeddings and residual codes that distinguish
descriptors within each cell.
The encoder produces $z_d=\enc_\nu(x_d)$, which the two layers
discretize:
\begin{itemize}
	\item  \emph{Coarse-cell layer:} The shared codebook is $U_\nu=\{\mu_c\}_{c=1}^{M_{\mathrm{code}}}$, where $M_{\mathrm{code}}$ is the number of centroid vectors and $\mu_c$ is the centroid of
	coarse cell $c$. Sorting the cell IDs $c$ by cosine distance from $z_d$ to
	$\mu_c$ gives the descriptor's coarse-cell ranking. We write
	$Z_d=\toprho(z_d;U_\nu)$ for the set of
	the top-$\rho$ cell IDs, where $\rho$ is the descriptor's coarse-cell
	depth. Retaining more than one cell ($\rho=2$ or $3$
	in practice) covers descriptors near a cell boundary. For each selected
	cell ID $c\in Z_d$, the residual vector
	$r_{d,c}=z_d-\mu_c$ represents the descriptor's position relative to
	centroid $\mu_c$.

\item \emph{Residual-code layer:} Each of $J$ independently seeded
\emph{residual-code families} maps
$r_{d,c}$ to an $\ell$-bit residual code
$b_{d,c}^{(j)}=\mathrm{Code}_\nu^{(j)}(r_{d,c})$, $j=1,\dots,J$.
Our concrete instantiation uses sign projection: family $j$ contains
projection vectors $a_{j,r}$, $r=1,\dots,\ell$, and packs the signs of
$\langle a_{j,r},r_{d,c}\rangle$ into $b_{d,c}^{(j)}$.
\end{itemize}

Figure~\ref{fig:sketch} illustrates how the coarse-cell and residual-code
layers produce publication keys for a descriptor and probe keys for a
query. In the figure's concrete example, $Z_d=\{17,42\}$, so the coarse
layer retains cell IDs 17 and 42 for descriptor $d$. In the zoomed view
of cell 17, the residual $r_{d,17}=z_d-\mu_{17}$ has sign pattern $00$
under residual-code family $j=1$, giving $b_{d,17}^{(1)}=00$. A query
that also selects cell 17 and obtains code $00$ under this family later
derives the corresponding probe key that matches $d$'s publication key.

\begin{figure}[t]
\centering
\begin{tikzpicture}[scale=0.62,every node/.style={font=\scriptsize}]

\begin{scope}
\draw[black!30,rounded corners] (-2.6,-0.9) rectangle (4.0,3.6);

\filldraw[black!35] (-1.6,2.8) circle (2.2pt);
\node[black!45,above=0pt] at (-1.6,2.8) {$\mu_{9}$};
\draw[black!25,dashed] (3.28,0.60) -- (-1.6,2.8);

\filldraw[vizorange] (2.6,0.8) circle (2.2pt);
\node[vizorange!80!black,above=0pt] at (2.6,0.8) {$\mu_{42}$};
\draw[vizorange!70,thick] (3.28,0.60) -- (2.6,0.8);

\filldraw[vizblue] (3.2,0.4) circle (2.2pt);
\node[vizblue!80!black,below=0pt] at (3.2,0.4) {$\mu_{17}$};
\draw[vizblue!70,thick] (3.28,0.60) -- (3.2,0.4);

\filldraw[black] (3.28,0.60) circle (2.4pt);
\node[black,above right=0pt] at (3.28,0.60) {$z_d$};

\node[align=left,black!55,below] at (0.7,-1.05)
  {$Z_d=\toprho(z_d;U_\nu)=\{17,42\}$, $\rho=2$};
\end{scope}

\begin{scope}[xshift=6.7cm]
\draw[black!30,rounded corners] (-1.8,-1.8) rectangle (1.8,1.8);
\node[black!55,above] at (0,1.95) {cell 17 (local, zoomed)};

\draw[black!35] (-1.6,0) -- (1.6,0);
\draw[black!35] (0,-1.6) -- (0,1.6);
\node[black!45] at (1.15,1.15) {$b=00$};
\node[black!45] at (-1.15,1.15) {$b=01$};
\node[black!45] at (-1.15,-1.15) {$b=11$};
\node[black!45] at (1.15,-1.15) {$b=10$};

\filldraw[vizblue] (0,0) circle (2pt);
\node[vizblue!80!black,below left=1pt,xshift=1.5pt,yshift=1.5pt]
  at (0,0) {$\mu_{17}$};
\draw[-{Latex[length=1.8mm]},vizblue,thick] (0,0) -- (0.24,0.6);
\filldraw[black] (0.24,0.6) circle (2pt);
\node[black,right=0pt] at (0.24,0.6) {$r_{d,17}$};
\node[vizblue!70!black,align=center,below] at (0,-1.95)
  {$b_{d,17}^{(1)}=\mathsf{00}$};
\end{scope}
\end{tikzpicture}
\caption{Left: $z_d$ selects top-$\rho{=}2$ coarse-cell IDs
$Z_d=\{17,42\}$ (blue/orange), while a farther centroid $\mu_9$
(gray) is not selected. Right: inside cell 17, the residual $r_{d,17}
= z_d-\mu_{17}$ is discretized by a sign-projection residual-code family
into a residual code; each residual-code family uses independent projection vectors,
giving residuals separated by one projection boundary additional
opportunities to match.}
\label{fig:sketch}
\end{figure}
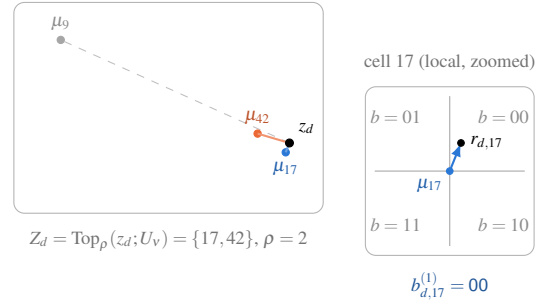

Coarse-cell overlap provides coverage, while
residual-code agreement provides discrimination within a shared cell.
Multiple residual-code families mitigate projection-boundary effects:
residual vectors separated by one family's boundary can receive
the same code under another family.
\nsclarify{Increasing $\rho$ or $J$ adds opportunities for shared keys
and increases publication fan-out; coverage depends on the budget.}
Appendix~\ref{app:hit-analysis} derives the
precision-key hit-probability model; Section~\ref{sec:eval:theory}
tests its residual-code component.

\subsection{Exact Keys and Bounded Publication}
\label{sec:index:keygen}
For each $c\in Z_d$, the provider derives one recall key and one
precision key per residual-code family. For namespace label $n$, cell
$c$, family $j$, and residual code $b$, the constructors are
\begin{equation}
\kappa_\nu^R(n,c)=H(\text{``R''}\|\nu\|n\|c),
\label{eq:recall-key}
\end{equation}
\begin{equation}
\kappa_{\nu,j}^P(n,c,b)=H(\text{``P''}\|\nu\|n\|c\|b\|j).
\label{eq:precision-key}
\end{equation}
The tags distinguish recall and precision keys; $j$ separates residual
codes from different families. Including $n$ enforces exact
compatibility boundaries. The provider substitutes $n(d)$ and
$b_{d,c}^{(j)}$ to obtain
\begin{equation}
K_{\mathrm{pub}}(d)=\bigcup_{c\in Z_d}\Bigl(
\{\kappa_\nu^R(n(d),c)\}\cup
\{\kappa_{\nu,j}^P(n(d),c,b_{d,c}^{(j)})\}_{j=1}^{J}\Bigr).
\label{eq:publication-key-set}
\end{equation}
Each selected cell contributes at most $1+J$ keys, yielding
\begin{equation}
|K_{\mathrm{pub}}(d)|\leq\rho(1+J).
\label{eq:publication-bound}
\end{equation}
This bound is independent of corpus size. Certification binds the
descriptor to this set before publication; see
Section~\ref{sec:poison:anchor}.

\subsection{Budgeted Probe Sequence}
\label{sec:index:probe}
The requester computes $z_q=\enc_\nu(x_q)$ and a ranked cell sequence
$Z_q^{\mathrm{ord}}$; \nsclarify{its first $\rho_q$ IDs form the primary-cell set $Z_q$.} For each
cell, it computes residual codes using the provider's
construction \nsclarify{and substitutes $n\in\mathcal N_q$ into the
key constructors.}

Probes follow three stages. $P_1(q)$ contains exact precision
keys for primary cells. \nsclarify{$P_2(q)$ probes codes at Hamming distances
one through $r_H$ from the query's residual codes in $Z_q$,} then exact precision
keys for secondary cells ranked $\rho_q+1$ through
$\rho_{\mathrm{ext}}$. Here $r_H=0$ disables neighboring-code probes,
and $\rho_{\mathrm{ext}}$ bounds the considered cell ranks.
\nsclarify{$P_3(q)$ contains recall keys for $Z_q$ in cell-rank order.} Within each stage,
the requester interleaves allowed namespace labels. The logical sequence is
\begin{equation}
\mathcal P(q)=\mathrm{Dedup}\bigl(P_1(q)\Vert P_2(q)\Vert P_3(q)\bigr),
\label{eq:logical-probe-sequence}
\end{equation}
where $\Vert$ concatenates sequences and \textsc{Dedup} retains each
key's first occurrence. Appendix~\ref{app:probe-order} gives the
deterministic ordering rules.

\nsclarify{With one entry point, unsharded keys, and no lineage-resolution
lookups, each probe costs one lookup, giving the budgeted key set}
\begin{equation}
K_{\mathrm{qry}}(q;L_q)=\mathrm{set}(\pref_{L_q}(\mathcal P(q))),
\label{eq:budgeted-probe-key-set}
\end{equation}
where $\pref_L$ takes a sequence's first $L$ elements. Shared
publication and probe keys expose candidate postings. The order uses
selective residual matches before broader recall lists; a smaller
\nsclarify{$L_q$ stops the sequence earlier. Lineage resolution,
repeated entry-point probes, and manifest or shard lookups draw from
the same remaining budget, leaving fewer lookups for subsequent probes,
as detailed in Section~\ref{sec:poison}.}
\endgroup

\section{\nsrevision{Certified Publication and Lookup}}
\label{sec:poison}
\begingroup\color{black}
Certification makes the publication bound enforceable at storage:
the committee derives the allowed keys, and posting-list services
verify each posting's membership. Requesters apply the same checks
to returned records before admitting candidates.

\subsection{Threats and Certification Assumptions}
\label{sec:poison:threat}
\label{sec:poison:assumptions}
An adversary may control providers, routing peers, and up to $f_A$
members of an anchor committee. We consider publication at
unauthorized keys, posting-list stuffing, suppression of valid
postings, and replay of stale versions. Certification establishes
consistency between a committed descriptor and its publication keys;
capability truthfulness and execution behavior are outside discovery.
For example, a membership certificate for the clinic's appointment
service does not establish clinical credentials, slot availability,
or the personal assistant's permission to invoke that service. Namespace
matching checks declared compatibility; it does not attest runtime
policy enforcement.

The deployment authenticates public keys and configuration $\nu$
and assigns each lineage to an $n_A$-member committee with signing
threshold $t_A$. With unforgeable signatures and collision-resistant
hashes, $f_A<t_A$ ensures that a valid committee signature includes an
honest member that recomputes the publication keys. Availability also
requires $n_A-f_A\geq t_A$ and progress of the signing service.

The committee serializes registrations, renewals, updates, and
revocations, rejects incompatible states at one epoch, and preserves
lineage history across committee changes. This state-serialization
service is an additional deployment assumption. Admission and
committee-selection policies constrain creation of provider identities
and grinding for favorable committees; quotas apply to admitted providers.

\subsection{Certified Publication}
\label{sec:poison:anchor}
\label{sec:poison:properties}
The provider submits the complete descriptor and a signed registration
request to the committee assigned to lineage $\iota_d$. The stable
anchor key $\kappa_A(\iota_d)=H(\text{``A''}\|\iota_d)$ locates that
lineage's certified state. The committee verifies the provider
signature, descriptor commitment, and requested lineage transition.
It recomputes the namespace, embedding, and publication-key set under
$\nu$, then commits to the sorted keys with a Merkle tree
~\cite{merkle1988signature}:
\begin{equation}
\field{root}_d=\merkle(\mathrm{sorted}(K_{\mathrm{pub}}(d))).
\label{eq:publication-root}
\end{equation}
The membership certificate $\Cert_d$ binds this root to $C_d$, the
provider key, namespace, configuration, lineage, epoch, predecessor,
live or tombstone state, and lease. The committee signs it as $\Sigma_d$.

\nsclarify{For each target key $\kappa$, the provider constructs a posting
body $m_{d,\kappa}$ that binds $\kappa$, descriptor commitment $C_d$,
retrieval address, lease, and certificate hash. The body includes
the provider's signature. It publishes}
\begin{equation}
\mathrm{Posting}(d,\kappa)=
\langle m_{d,\kappa},\Cert_d,\Sigma_d,\pi_{d,\kappa}\rangle,
\label{eq:certified-posting}
\end{equation}
where $\pi_{d,\kappa}$ proves the key's inclusion under
$\field{root}_d$. The signed certificate supplies the provider key,
so verification does not require fetching the complete descriptor.
Appendix~\ref{app:certified-records} specifies the record fields.

\subsection{Posting and Candidate Verification}
\label{sec:poison:accept}
\label{sec:poison:narrow}
\label{sec:index:narrow}
A posting-list service checks that the received key matches the
posting body and that the body's commitment, provider key, namespace,
lineage, epoch, lease, and certificate hash agree with $\Cert_d$.
It verifies the provider signature, the assigned committee's signature
and threshold, the supported configuration, and the Merkle proof.
These checks restrict an accepted posting to $K_{\mathrm{pub}}(d)$.
Keeping one active posting per $(C_d,\kappa)$ prevents repeated
publication from enlarging the list; together with
Equation~\eqref{eq:publication-bound}, it enforces the logical fan-out bound.
Optional admission-layer quotas bound active states per provider and
\nsclarify{namespace, as specified in Appendix~\ref{app:posting-verification}.}

Both the service and requester authenticate an incoming certificate
before adding it to observed lineage state. They select the highest
observed epoch, then require a consistent live state and an unexpired
lease. Epoch and revocation evidence survives expiry; an expired
highest epoch does not restore an older version. Missing predecessor
links or conflicting states trigger resolution at the stable anchor
\nsclarify{key, using the requester's remaining lookup budget.
Probing and resolution stop when $L_q$ is exhausted; unresolved
lineages supply no candidates.}

The requester verifies returned postings using its own time and
observations. It merges accepted postings by $C_d$, fetches each
complete descriptor once, and checks its commitment, provider key,
namespace, and lineage against the certificate. \nsclarify{After checking
$n(d)\in\mathcal N_q$ and resolving versions, it computes
$z_d=\enc_\nu(x_d)$ from the verified descriptor and ranks by the base score:}
\begin{equation}
s_{\mathrm{base}}(d\mid q)=\mathrm{sim}(z_q,z_d).
\label{eq:base-score}
\end{equation}
\nsclarify{Here $\mathrm{sim}$ denotes cosine similarity, and $\cand{q}$
contains up to $k_q$ highest-scoring eligible candidates.}
Certificate consistency and freshness determine eligibility before
ranking. A rejected posting alone does not attribute misconduct to
its provider: \nsclarify{a peer can alter proofs or replay old postings.}

\subsection{Maintenance and Additional Lookup Paths}
\label{sec:poison:maintenance}
Renewal extends the lease of a certified live state at the same epoch;
an update advances the epoch, and a tombstone terminates the lineage.
An update can also change the publication-key set, so requesters
probing old keys need a way to discover the successor. During a
bounded migration window, the provider publishes the successor at
its new keys and leaves Move hints at the old keys. Each hint carries
the successor's membership certificate and committee signature.
The requester verifies the hint and resolves the lineage through the
stable anchor key before applying the candidate checks in
Section~\ref{sec:poison:narrow}.

Read-repair propagates verified lineage state to posting-list services
that return older versions. The requester sends the newer membership
certificate, committee signature, and necessary predecessor evidence;
the recipient verifies the signatures and lineage transitions before
updating its state. Requester DHT lookups for migration and repair
consume the remaining $L_q$; direct repair messages and recipient
background fetches are accounted for as maintenance traffic.
Appendix~\ref{sec:freshness} specifies the transitions and verification
rules.

Physical sharding redistributes a posting list while preserving its
logical key and membership checks; manifest and shard lookups consume
\nsclarify{$L_q$, as detailed in Appendix~\ref{sec:scaling}.}
A requester can also repeat selected
keys across entry points to obtain observations through different
paths. Each repetition consumes budget, trading probe depth for
additional observations. \nsclarify{Appendix~\ref{app:multi-view} defines
the multi-view schedule and optional coverage-based ranking.}
\endgroup

\section{\nsrevision{Implementation}}
\label{sec:impl}
\begingroup\color{black}

\nsclarify{We implement \system's storage and lookup paths in Go.
The prototype combines a DHT overlay with a replicated posting-list
service. Providers publish certified postings to responsible peers;
requesters locate those peers, retrieve and verify postings, and merge
the results.}

\subsection{DHT Routing and Posting-List Service}
\nsclarify{The prototype uses Go 1.25.7 with libp2p v0.49.0 for peer
connections~\cite{benet2014ipfs} and go-libp2p-kad-dht v0.42.1 for
exact-key routing~\cite{maymounkov2002kademlia}. Many providers can
publish at one key, so the posting-list service supplies multi-record
storage beyond libp2p's single-value API\@. Replicas acknowledge postings
after verification, journaling, and indexing by logical key and
descriptor commitment.}

\subsection{Certified Records and Verification}
The wire format uses canonical CBOR~\cite{bormann2020cbor}. Providers
sign posting bodies with Ed25519~\cite{josefsson2017eddsa}; committee
signatures use CIRCL's BLS12-381 implementation
~\cite{boneh2004bls,boneh2003aggregate}.
\nsclarify{Each member signs a message derived from the common certificate
digest and its member index; a bitmap identifies the signers.
A shared verifier retrieves their public keys and threshold from the
committee registry. At storage admission and requester retrieval, it
checks the target key, consistency with the membership certificate,
provider-key binding, lease, both signatures, and Merkle inclusion proof.}

\subsection{Paging, Replica Reads, and Certificate Caching}
\nsclarify{Paging bounds each response's size. Replicas return postings
in commitment order with a continuation cursor, a generation counter
for list changes, and a total record count.}
The requester checks ordering, cursor progress, duplicate commitments,
generation consistency, and the final count before accepting the
replica's response. \nsclarify{It reads replicas concurrently, waits
for every attempt to finish, requires the read quorum, and merges
valid postings by descriptor commitment. This completion policy lets
a slow replica delay the lookup even after a quorum responds.}

\nsclarify{A bounded FIFO cache reuses committee-signature verification
across keys. Its key hashes the complete encoded membership certificate
and committee signature, including the signer bitmap and aggregate
signature. Hits skip committee-signature verification but retain all
posting-specific checks, including the current lease check. Invalid
certificates are never cached; changing the signature changes the cache key.}

\subsection{Execution and Validation Scope}
\nsclarify{Before network execution, we freeze embeddings, publication
keys, probe sequences, and expected candidate commitments, then
materialize and publish certified postings. The driver replays these
fixed plans on the running overlay, timing routing, paged reads,
verification, and candidate reconstruction. Encoding, committee issuance,
complete-descriptor retrieval, and capability execution remain outside
this interval. The prototype exercises the common index
and posting-verification path using API-derived descriptors. It does
not implement an appointment application or adapters for conversational
agent endpoints and arbitrary service APIs.
The store retains observed lineage state to reject local
rollback; distributed renewal, migration, and shard-layout mechanisms
are evaluated separately, as detailed in Appendices~\ref{sec:freshness}
and~\ref{sec:scaling}.}
\endgroup

\section{\nsrevision{Evaluation}}
\label{sec:eval}
\begingroup\color{black}
\nsclarify{We first evaluate retrieval effectiveness and index efficiency,
measuring recall, candidate exposure, publication fan-out, and lookup
work in Section~\ref{sec:eval:rq1}. We then test whether lookup savings
translate into lower completion time over the Internet, and examine
scheduling, verification, and resource costs in
Section~\ref{sec:eval:network}. Finally, Section~\ref{sec:eval:rq4}
validates candidate reconstruction and protocol behavior under faults,
unauthorized publication, and descriptor updates.
Section~\ref{sec:discussion:limitations} discusses the evaluation's
scope and deployment limitations.}

\subsection{Experimental Setup}
\label{sec:eval:setup}

The source corpus is ToolBench/RapidAPI metadata
~\cite{qin2023toolllm}: 39{,}529 API descriptors after filtering
near-empty descriptions, spanning 1{,}433 namespaces identified by
$n(d)=(\tau_d,\sigma_d,\psi_d)$. Query traffic comprises 8{,}087
ToolBench retrieval queries. \nsrevision{Its relevance judgments were generated by an LLM that
reconstructed task intents from known relevant APIs.}

These API descriptions exercise the common capability
index and its network lookup path. The evaluation does not measure
conversational interaction with agent endpoints or downstream service
execution. The appointment scenario is an illustrative use case,
not a deployed application in this workload.

We encode descriptor and query text with the 384-dimensional
\texttt{all-MiniLM-L6-v2} checkpoint
~\cite{reimers2019sentencebert}, using Sentence Transformers 5.6.1,
normalized \texttt{float32} vectors, and the same
canonical input template for descriptor and query encoding. The coarse layer
uses $M_{\mathrm{code}}=16$ centroids trained for 25 k-means iterations. Each of three
reported seeds regenerates the codebook and residual-code families and
is carried in the versioned semantic-index configuration, so providers and requesters use
the same authenticated semantic-index configuration.

Because roughly a third of query-reachable namespaces are too sparse
to expose selectivity, the larger stress corpus adds 41{,}295
query-conditioned, semantically adjacent LLM variants, capped at 20
per real seed. Hard-constraint fields are copied from the source
descriptor and every generated descriptor retains provenance. \nsrevision{This controlled density augmentation yields
80{,}824 descriptors.} Section~\ref{sec:eval:rq1} first reports
the real-only corpus, while the parameter, baseline, and robustness
sweeps use the combined stress corpus unless noted. Namespace-scoped
exact cosine top-$k$ over the same encoder is our index-fidelity oracle;
ToolBench relevance labels provide a non-circular but
noisier measurement.

\subsection{\nsclarify{Retrieval Effectiveness and Index Efficiency}}
\label{sec:eval:rq1}

On the real-only corpus, before density augmentation, one calibrated
configuration reaches $\mathrm{recall}@10=0.955$ against exact cosine
neighbors at physical lookup budget $L_q=32$, and $0.947$ against
ToolBench's dataset labels. \nsrevision{The two measurements assess agreement with the
encoder's nearest neighbors and with the dataset's relevance labels.}

The reported configurations tie query and publication coarse depth,
$\rho_q=\rho$, and list their shared value as $\rho$. We obtain a
sharper view than a single recall number by sweeping
\nsrevision{$(\rho,J,r_H)\in\{1,2,3,4\}\times\{1,2,3,4\}\times\{0,1,2\}$}
(48 configurations, with $\ell$ and $\rho_{\mathrm{ext}}$ derived from $\rho$
rather than swept independently, since an uncalibrated $\ell$ simply
produces
degenerate key populations rather than new information) on the combined
corpus. We report three selectivity-first operating points in
Table~\ref{tab:e2-operating-points}.

\begin{table}[t]
\centering\small
\begin{tabular}{@{}rlrrr@{}}
\toprule
recall target & $(\rho,J,\ell,r_H,L_q)$ & recall & exposure & fan-out \\
\midrule
0.90 & $(3,4,3,1,32)$ & 0.902 & 62.6\% & 15 \\
0.95 & $(2,3,3,1,128)$ & 0.950 & 72.8\% & 8 \\
0.97 & $(3,2,3,2,64)$ & 0.970 & 80.1\% & 9 \\
\bottomrule
\end{tabular}
\caption{Selectivity-first operating points from the full parameter
sweep. ``exposure'' is the fraction of the query's reachable
candidate population actually exposed at that budget---lower is more
selective at the same recall.}
\label{tab:e2-operating-points}
\end{table}

\begin{figure}[t]
\centering
\includegraphics[width=0.92\linewidth]{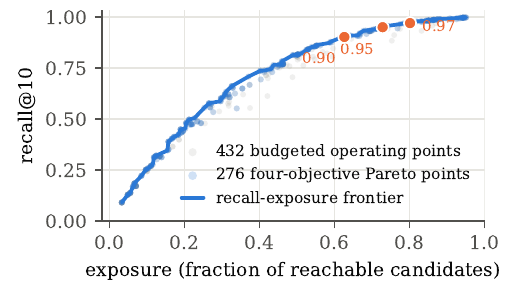}
\caption{Recall versus exposure for 432 budgeted
operating points from 48 $(\rho,J,r_H)$ configurations on the combined
corpus. Blue points are the 276 operating points retained by the
four-objective Pareto filter; the curve traces the two-dimensional
recall--exposure frontier. The three marked operating points
correspond to Table~\ref{tab:e2-operating-points}.}
\label{fig:e2-pareto}
\end{figure}

We apply the four-objective Pareto filter to all 432 operating points;
Figure~\ref{fig:e2-pareto} presents the full and retained sets.
The operating points expose the interaction between sketch depth
and the query budget. Increasing $\rho$, $J$, or $r_H$ at fixed
$L_q$ lengthens the precision-probe prefix and can delay the
coarse-recall fallback enough to reduce recall. The Pareto filter
therefore retains tradeoffs among recall, exposure, lookup count,
and publication fan-out. At recall at least $0.90$, the smallest
observed exposure is $62.6\%$ of the reachable population. Raising
recall to the sweep's maximum of $0.9976$ requires $95.0\%$ exposure.

Two additional codebook seeds preserve the ordering of the three
representative points: their recall ranges are $0.902$--$0.919$,
$0.950$--$0.960$, and $0.970$--$0.978$, respectively. Exposure varies
by 3--5 percentage points across seeds. Appendix~\ref{app:retrieval-details}
reports the query-bootstrap intervals and the baseline selection
details.

\paragraph{\nsclarify{Publication and lookup cost.}}
\label{sec:eval:rq2}

We compare \system with LSH-on-DHT over the same namespaces and
corpus, selecting configurations at matched recall. The LSH sweep
combines signature widths $w_{\mathrm{LSH}}\in\{4,6,8\}$, table
counts $T\in\{1,2,4,8,16\}$, and Hamming radii
$r_H\in\{0,1,2\}$: 45 configurations at nine budgets, yielding
405 operating points. For each system, we linearly interpolate
each configuration's budget curve to the target recall
and select the configuration with the smallest exposure.

Centralized HNSW~\cite{malkov2020hnsw} and exact FlatIP provide local
retrieval references. We sweep HNSW graph degree
$M_{\mathrm{HNSW}}\in\{8,16,32\}$ and search breadth
$\mathrm{efSearch}\in\{8,\ldots,256\}$. We also evaluate exact-key
hashing of capability names and an inverted BM25
index~\cite{robertson2009bm25} within the same namespaces.

At recall $0.90$, \system and LSH expose $62.1\%$ and $61.6\%$
of the reachable candidate population, while issuing 33.7 and
173.2 exact-key lookups on average. At recall $0.95$, both expose
$72.7\%$, with 32.9 versus 253.5 lookups. The corresponding lookup
reductions are $5.1\times$ and $7.7\times$; publication fan-out at
recall $0.95$ is ten versus sixteen. Thus, the key construction
reaches comparable candidate populations through fewer posting-list
accesses. Appendix~\ref{app:retrieval-details} gives the complete
matched-recall table and explains why its interpolated fan-out-ten
point differs from the directly selected fan-out-eight point in
Table~\ref{tab:e2-operating-points}.
Figure~\ref{fig:baseline} shows how this lookup advantage varies
across the four recall targets. The gap narrows at $0.97$, where
the selected LSH signature width and probe radius change.

\begin{figure}[t]
\centering
\includegraphics[width=0.85\linewidth]{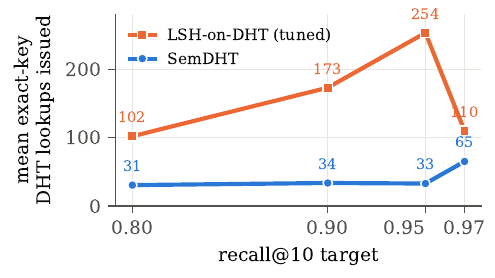}
\caption{Mean exact-key DHT lookups issued to reach a given recall
target, \system versus the best-tuned LSH-on-DHT configuration at that
\nsclarify{target, using the data presented in Table~\ref{tab:baseline-headline}.} \system is
cheaper at all four targets; the gap is largest at recall $0.95$ and
narrows at $0.97$ because the matched LSH optimum switches from
the hash-width/radius pair $(w_{\mathrm{LSH}},r_H)=(8,2)$ to $(6,1)$.}
\label{fig:baseline}
\end{figure}

HNSW reaches recall $0.985$ at $M_{\mathrm{HNSW}}=16$ and
$\mathrm{efSearch}=16$, while exact FlatIP reaches $1.0$. Their
reported times measure local batch CPU work. With verbatim API-name
matching, exact-key hashing reaches recall $0.065$. BM25's top-10
ranking reaches about $0.398$ recall with mean
publication fan-out $26.2$ ($p95=39$); its candidate population
contains enough relevant items for oracle recall $0.953$. This
difference separates candidate coverage from ranking quality.
Two LSH seeds give recall $0.94$--$0.96$ and preserve
the lookup-cost ordering at the tested targets.

\paragraph{\nsclarify{Residual-code hit model.}}
\label{sec:eval:theory}

Multiple residual-code families give a query--descriptor pair
\nsclarify{additional opportunities for a precision-key hit,
as described in Section~\ref{sec:index:sketch}.} We test how well the angle-based
prediction and approximate-independence assumption in
Appendix~\ref{sec:index:theory} describe measured residual-code hits.
We use query--descriptor residual pairs from the combined corpus
across three codebook seeds: $80{,}858$ namespace-correct nearest
pairs and $80{,}830$ random reachable negative pairs.

Figure~\ref{fig:a3-theory} compares measured and predicted single-family
hit probabilities. For the residual-code
component with $J=4$, the observation-weighted mean absolute error
is $0.0022$. The maximum normalized conditional covariance
between residual-code families is $0.0131$, below the prespecified
$0.05$ threshold for approximate independence.

\begin{figure}[t]
\centering
\includegraphics[width=0.92\linewidth]{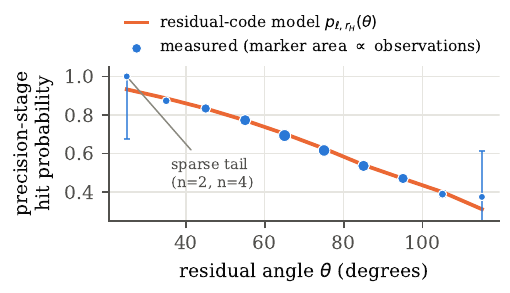}
\caption{Measured single-family hit probability versus residual
angle $\theta$ at a representative calibrated configuration
($\rho{=}2,\ell{=}3,r_H{=}1$), against the closed-form
$p_{\ell,r_H}(\theta)$ prediction from Appendix~\ref{sec:index:theory}. Marker
area is proportional to the number of real observations in that bin;
error bars are 95\% confidence intervals. The fit is close across the
well-populated $30^\circ$--$100^\circ$ range and only visibly departs
in the sparsely observed tails.}
\label{fig:a3-theory}
\end{figure}

At a different
configuration ($\rho=4,\ell=4,r_H=1,J=4$), the sparse $100^\circ$--$110^\circ$
residual-angle bin shows a measured hit rate of $0.581$ against a
predicted $0.627$, an error of $-0.0461$. This bin covers
$0.363\%$ of that configuration's observations. The aggregate error
and conditional covariance support the residual-code component of the
precision-key hit-probability model over the bulk of the observed
distribution, with a narrow tail deviation consistent with anisotropy
in the embeddings.

\subsection{\nsclarify{Network Performance and System Overhead}}
\label{sec:eval:network}
\label{sec:eval:rq3}
\begingroup\color{black}
\paragraph{Deployment and workload.}
\nsclarify{We evaluate whether fewer logical key accesses translate into
lower lookup completion time over the Internet. We deploy two 200-peer
overlays on Alibaba Cloud with the placements and machine specifications
in Table~\ref{tab:tierc-deployment}.} The requester runs on a dedicated
VM and replays queries sequentially against both deployments. A
separate 4-vCPU, 8-GB controller VM in Frankfurt manages deployment
and artifact collection. Each overlay uses three replicas,
write and read quorums of two, pages of at most 64 postings, and at
most 32 concurrent logical lookups. Lookup and RPC timeouts are
60\,s. Replica reads use the wait-all completion policy described in
Section~\ref{sec:impl}. Both \system and LSH use the same certified
posting format, replica policy, and posting-list service.

\begin{table}[t]
\color{black}
\centering\small
\setlength{\tabcolsep}{3pt}
\begin{tabular}{@{}llrrrr@{}}
\toprule
Deployment & Region & VMs & Peers/VM & vCPU & RAM \\
\midrule
Same-region & Frankfurt & 8 & 25 & 4 & 8\,GB \\
Cross-region & Eight regions & 8 & 25 & 4 & 8\,GB \\
Requester & Frankfurt & 1 & -- & 8 & 16\,GB \\
\bottomrule
\end{tabular}
\caption{Network-replay deployment. Each overlay has eight virtual
machines (VMs), each hosting 25 peer processes. The cross-region
deployment places one VM in each of Frankfurt, Tokyo, Singapore,
Hong Kong, Virginia, Silicon Valley, S\~ao Paulo, and Dubai.
Peers/VM counts overlay service processes; hardware specifications
are per VM\@. All machines use Intel Xeon Platinum CPUs and 64-bit
Ubuntu 24.04, with 10\,Mb/s of configured bandwidth per machine.}
\label{tab:tierc-deployment}
\end{table}

\nsclarify{The replay uses posting lists and probe sequences materialized
from the frozen 500-query workload. Setup publishes the postings and
reads back the planned keys, including empty keys, to verify their
contents before timing.}
The executable operating points use probe-key budgets of 64 for
\system and 256 for LSH\@. Their candidate-exposure recall on the
8{,}087-query source workload is 0.95198 and 0.95185, respectively;
\nsclarify{these configurations are fixed before network execution.}
The replay measures system cost and checks candidates against each
plan's expected commitments. We report 299 complete queries from
this frozen workload.
Each query runs under all 16 combinations of system, region placement,
parallel or stage-barrier scheduling, and cold or warm certificate
cache. Query order is seeded, and a balanced Latin square interleaves
the conditions.

\paragraph{Scheduling and measurement.}
\nsclarify{For each query and system, both schedulers execute every key
in the same budgeted probe sequence.}
The parallel scheduler lets keys from all stages compete for the 32
lookup slots. The stage-barrier scheduler permits concurrency within
a stage but admits the next stage only after every lookup in the
current stage finishes. The query budget determines the key
set before execution; retrieved candidates do not trigger early stopping.
\nsclarify{We time the lookup path described in Section~\ref{sec:impl},
using one batch for parallel scheduling and summing stage-batch times
for stage-barrier scheduling. Cache prewarming precedes this interval.}
We preserve the frozen sample weights
and resample whole queries, retaining all 16 conditions together, for
2{,}000 bootstrap repetitions. \nsclarify{We report weighted means
and P95 with 95\% intervals.} \nsclarify{Appendix~\ref{app:tierc-replay}
details the conditions, sampling, and late-run CPU instrumentation.}

\begin{figure}[t]
\color{black}
\centering
\includegraphics[width=\linewidth]{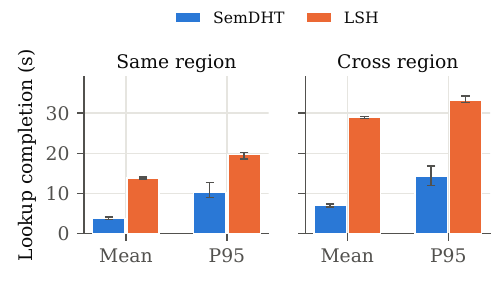}
\caption{Measured lookup completion with parallel scheduling and a
cold certificate cache. Bars show the weighted mean and P95; error bars
are 95\% query-cluster bootstrap intervals. Sample: 299 complete queries.}
\label{fig:tierc-latency}
\end{figure}

\paragraph{Completion time and lookup work.}
\nsclarify{Figure~\ref{fig:tierc-latency} presents completion times with
parallel scheduling and a cold certificate cache.} \system completes
in 3.819\,s on average within a region and 7.038\,s across regions,
compared with 13.897\,s and 28.938\,s for LSH\@. These correspond to
$3.64\times$ and $4.11\times$ lower mean completion times. P95 is
10.289\,s versus 19.721\,s within a region, and 14.178\,s versus
33.318\,s across regions. The paired mean reductions have 95\% intervals
of [9.839, 10.294]\,s and [21.438, 22.313]\,s, respectively.

\begin{table}[t]
\color{black}
\centering\small
\begin{tabular}{@{}llrrr@{}}
\toprule
Placement & System & Lookups & RPCs & MiB \\
\midrule
Same & \system & 32.59 & 130.91 & 2.913 \\
Same & LSH & 256.00 & 777.85 & 2.945 \\
Cross & \system & 32.59 & 130.91 & 2.913 \\
Cross & LSH & 256.00 & 777.85 & 2.945 \\
\bottomrule
\end{tabular}

\caption{Weighted mean work per query, parallel scheduling and cold
certificate cache. Lookups count logical keys; RPCs include replica
reads and pagination; MiB counts requester application-message frames.}
\label{tab:tierc-overhead}
\end{table}

\nsclarify{Table~\ref{tab:tierc-overhead} reports per-query work.
The probe-key budgets are upper bounds;
\system issues 32.59 logical lookups on average, versus 256 for LSH,
a reduction of 87.3\%. Replica reads and pagination produce 130.91
and 777.85 application RPCs, respectively, while message volume remains
close at 2.913 versus 2.945\,MiB. The systems transfer comparable
posting volumes, while \system accesses fewer keys. Each
key requires routing and replica reads, so fewer keys reduce the remote
operations that must complete under the same concurrency limit.}
Bytes count requester application frames, excluding connection setup,
other DHT routing traffic, transport retransmissions, and IP/TCP headers.

\begin{table*}[t]
\color{black}
\centering\small
\setlength{\tabcolsep}{4pt}
\begin{tabular}{@{}lllrr@{}}
\toprule
Placement & System & Cache & Mean difference [95\% CI] (s) & P95 difference [95\% CI] (s) \\
\midrule
Same & \system & Cold & 0.157 [0.075, 0.235] & 1.118 [-0.948, 2.124] \\
Same & \system & Warm & -0.091 [-0.139, -0.045] & -0.127 [-0.246, 0.291] \\
Same & LSH & Cold & -0.067 [-0.175, 0.044] & 0.014 [-0.618, 0.742] \\
Same & LSH & Warm & 0.000 [-0.118, 0.116] & -0.064 [-0.521, 0.809] \\
Cross & \system & Cold & 2.679 [2.540, 2.834] & 4.133 [2.315, 5.101] \\
Cross & \system & Warm & 2.259 [2.114, 2.409] & 0.854 [-0.478, 3.830] \\
Cross & LSH & Cold & 3.224 [2.947, 3.517] & 2.973 [1.794, 4.392] \\
Cross & LSH & Warm & 3.085 [2.833, 3.339] & 2.835 [2.299, 4.018] \\
\bottomrule
\end{tabular}

\caption{Scheduling contrasts over the same 299 query clusters:
stage-barrier minus parallel completion time. Positive values indicate
longer completion with barriers. P95 contrasts subtract the two
weighted P95 estimates; they are not percentiles of per-query
differences. Intervals retain query pairing and frozen sample weights.}
\label{tab:tierc-scheduling}
\end{table*}

\paragraph{Scheduling and completion.}
\nsclarify{Table~\ref{tab:tierc-scheduling} compares scheduling policies
while holding each query's keys fixed, for both placements and cache
states.} Stage barriers have
a larger effect across regions: with cold caches, they increase mean
completion time by 2.679\,s for \system and 3.224\,s for LSH\@.
Within-region changes are small; the cold-cache LSH interval includes
zero. The cross-region mean increases persist after certificate
prewarming, while within-region changes stay below 0.1\,s in magnitude.

Cold-cache P95 shows the same placement dependence: stage barriers
increase cross-region P95 for both systems, while both within-region
intervals include zero. With warm caches, the cross-region P95
interval remains positive for LSH but includes zero for \system.
\nsclarify{Under the wait-all replica policy, a lookup remains active
until its last replica attempt ends. A stage barrier extends this wait
to later stages; parallel scheduling lets their keys occupy free slots
while the slow lookup finishes. The benefit is larger across regions.
Both policies preserve \system's
lookup-count advantage.}
\nsclarify{With cold caches and stage barriers, \system achieves mean
completion speedups over LSH of} $3.48\times$ within a
region and $3.31\times$ across regions.

Across all 4{,}784 measured query-condition pairs, nine of
2{,}173{,}436 application RPCs report errors. Every logical lookup
still meets its read quorum, and every reconstructed candidate set
matches its expected set. All error-bearing observations remain in
the analysis. Appendix~\ref{app:tierc-replay} gives the error breakdown
and recording boundaries.
\endgroup

\paragraph{\nsclarify{Certification cost and cache behavior.}}
\label{sec:eval:certcost}
\begingroup\color{black}
We isolate the cost of posting validation with the protocol-v2 wire
format and CIRCL's portable BLS implementation on Windows/amd64.
The fixture has 16 authorized keys and five signers from a seven-member
committee. Each operation has ten independent benchmark repetitions;
Table~\ref{tab:tierc-validation} reports their means and Student-$t$
95\% intervals. Full validation takes 10.650\,ms per posting. Reusing
an already verified certificate reduces this to 48.7\,$\mu$s, a
$218.5\times$ reduction for this validation operation.
\nsclarify{Appendix~\ref{app:validation-fixture} details the fixture's
839-byte posting, including its provider signature, membership
certificate, committee signature, and Merkle inclusion proof.}

\begin{table}[t]
\color{black}
\centering\small
\begin{tabular}{@{}lrr@{}}
\toprule
Validation path & Mean (ms) & 95\% CI (ms) \\
\midrule
Cold certificate & 10.6496 & [10.5570, 10.7422] \\
Verified certificate cached & 0.0487 & [0.0471, 0.0504] \\
\bottomrule
\end{tabular}

\caption{Per-posting validation with five of seven committee signers
and a 16-key publication set. Cache hits retain posting-specific checks.}
\label{tab:tierc-validation}
\end{table}

The replay uses a separate 65{,}536-entry certificate cache for each
overlay, cleared before each condition. A warm condition first
preverifies the same query's certificates; this prelude is recorded
separately. For \system with parallel scheduling, warm-cache mean
completion is 3.009\,s within a region and 6.836\,s across regions,
compared with cold-cache means of 3.819\,s and 7.038\,s. Warm requests
have no certificate-cache misses. Their prelude itself averages
5.021\,s and 5.026\,s, respectively, so the warm results describe
reuse after verification. The remaining lookup work makes the query
speedup much smaller than the isolated validation speedup.
\endgroup

\paragraph{\nsclarify{Node CPU and memory use.}}
\label{sec:eval:resources}

We measure per-node resource use in a separate 400-node deployment,
with 375 nodes, the controller, and the requester on a local server,
and 25 nodes on one Hong Kong VM with the same 4-vCPU, 8-GB
configuration as the network-replay data VMs. The cloud nodes run
with \texttt{GOMAXPROCS=1}. We publish posting lists for
50 queries from the frozen workload and replay the same queries
under all eight combinations of system, scheduler, and certificate
cache state. Three rounds yield 24 complete batches and 1{,}200
query executions.

Each batch has a 300-second baseline followed by the query phase.
Every 10 seconds, we sample each cloud node's cgroup CPU counter
and process resident set size (RSS). CPU utilization divides consumed
core-seconds by elapsed time; 100\% denotes one fully occupied
logical core. Table~\ref{tab:tierc-resources} reports time-weighted
per-node means, averaged over the 25 nodes and then equally over
the three rounds. The query phase includes certificate prewarming,
waiting, and background DHT activity.

\begin{table}[t]
\centering\small
\setlength{\tabcolsep}{3pt}
\begin{tabular}{@{}llrrrr@{}}
\toprule
& & \multicolumn{2}{c}{CPU (\%)} & \multicolumn{2}{c}{RSS (MiB)} \\
\cmidrule(lr){3-4}\cmidrule(l){5-6}
Scheduler & Cache & \system & LSH & \system & LSH \\
\midrule
Parallel & Cold & 0.326 & 0.394 & 52.17 & 52.85 \\
Parallel & Warm & 0.291 & 0.440 & 52.16 & 53.00 \\
Stage-barrier & Cold & 0.336 & 0.410 & 52.50 & 52.68 \\
Stage-barrier & Warm & 0.294 & 0.431 & 51.68 & 52.89 \\
\bottomrule
\end{tabular}

\caption{Observed per-node resource use during the query phase
on the Hong Kong VM\@. Values average three rounds of 50 queries
per condition. Appendix~\ref{app:tierc-resources} reports baseline
values, round-to-round ranges, and query-validation results.}
\label{tab:tierc-resources}
\end{table}

\nsclarify{Across these conditions, \system averages 0.291--0.336\% of one
core per node, compared with 0.394--0.440\% for LSH\@. RSS is similar:
51.68--52.50\,MiB for \system and 52.68--53.00\,MiB for LSH\@.
Baseline CPU means are similar, at 0.398--0.416\%.}

\subsection{\nsclarify{Robustness and Protocol Validation}}
\label{sec:eval:rq4}

\nsclarify{We validate candidate reconstruction and protocol behavior under
faulty responses, unauthorized publication, and descriptor updates.
Implementation checks exercise the prototype, while controlled
simulations isolate publication membership and update behavior.
The appendices add multi-view lookup and storage experiments.}

\nsrevision{\paragraph{Implementation checks.}
Before network replay, a four-peer run with three replicas and read
and write quorums of two reconstructs the expected candidate sets for
all 500 queries under both systems. It publishes 398{,}196 postings
and reads all 103{,}545 planned keys, including empty lists. Separate
fault-injection checks cover missing or duplicate page contents,
generation changes, slow replicas, insufficient quorums, tampering
after cache prewarming, and cache eviction. Go and Python also agree
on the protocol-v2 canonical encodings and certificate digests.
The following experiments evaluate the design under controlled
publication attacks and state changes.}

\paragraph{Publication membership.}
\label{sec:eval:publication}
The 80{,}824-descriptor stress corpus contains 49{,}575 simulated
providers, grouped by distinct source provider identifier. We sample
malicious providers at fractions from $0\%$ to $50\%$ and have each
publish to a hot logical key outside its certified publication-key
set. With perfect routing and uncompromised certification, membership
verification rejects every injected posting. At $20\%$ malicious
providers, skipping membership verification produces $1.692\times$
query-weighted exposure; verification retains the honest recall of
$0.9500$. Figure~\ref{fig:e4} shows the full sweep: amplification
reaches $2.73\times$ without membership verification and remains
$1.000\times$ with it.

\begin{figure}[t]
\centering
\includegraphics[width=0.92\linewidth]{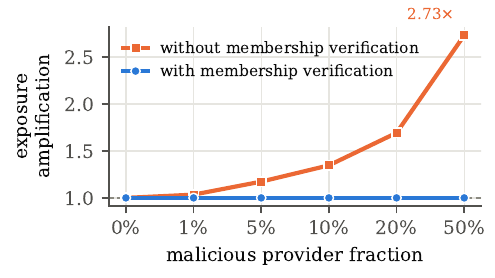}
\caption{Query-weighted candidate-acceptance amplification versus
malicious-provider fraction under perfect routing and uncompromised
certification. Membership verification keeps the measured ratio at
$1.000\times$ across the sweep.}
\label{fig:e4}
\end{figure}

\paragraph{Provider quotas.}
At $Q_{\mathrm{ns}}=128$, admission-layer quota accounting retains
$99.42\%$ of honest descriptors and reduces honest recall by $0.0023$.
Under a fixed attack budget, stuffing amplification rises from
$1.137\times$ with one attacking provider per namespace to
$2.081\times$ with 16 attacking providers. The quota limits one
admitted provider's publication volume, while additional identities
supply additional quota slots.

\paragraph{Descriptor updates.}
Controlled simulations evaluate migration and read-repair.
For 512 lineages with disjoint old and new publication-key sets,
a Move window covering the query-switch interval preserves
discoverability throughout the update. The model assumes successful
Move resolution and excludes its additional DHT lookup cost.
At replica turnover rate $0.10$ per normalized time unit,
read-repair lowers the p95 time for all eight replicas to hold the
new epoch from $17.96$ to $8.95$ units. Both arms include background
anti-entropy; repair adds an average of 1{,}538 direct messages per
512-lineage run over three seeds. A separate replay check rejects
older postings after a verified update or tombstone, before their
leases expire. Appendices~\ref{sec:eval:migration}
and~\ref{sec:eval:freshness} give the setups, window and turnover
sweeps, and maintenance costs.

\paragraph{Multi-view lookup and storage.}
Appendix~\ref{sec:eval:multiview} evaluates entry-path suppression.
At capture probability $f=0.20$, repeating the complete probe prefix
through two entries raises attacked recall from $0.7957$ to $0.9191$
with twice the lookup cost. At fixed cost, the shorter repeated
prefix reduces clean recall from $0.9500$ to $0.8553$.
Appendix~\ref{app:storage-sweeps} reports how sharding redistributes
response work under query skew and how cache revalidation affects
candidate acceptance. Physical-budget accounting, including repeated
entry-point probes and shard lookups, is evaluated in
Appendix~\ref{sec:scaling:cost}.

\endgroup

\Needspace{7\baselineskip}
\section{\nsrevision{Related Work}}
\label{sec:related}
\begingroup\color{black}

\paragraph{Agent-accessible services and capability discovery.}
A2A describes agent capabilities through Agent Cards,
with discovery through known locations, catalogs, or configuration
~\cite{a2a2025spec}. MCP defines tool descriptions and invocation
~\cite{mcp2025tools}, and its registry catalogs MCP servers
~\cite{mcpregistry}. ToolBench studies task-based API retrieval and
tool use~\cite{qin2023toolllm}. \system studies the index and network
costs of discovering advertised capabilities under supported interface
contracts, whether a provider exposes an agent endpoint or a service API.

Agent-network architectures develop capability announcements,
topology-independent naming, and semantic communication
~\cite{wang2026agenticp2p,rodriguez2026agentidentity,fleming2025layered,xu2026agentosi,zhang2026distributedagents}.
Guo et al.\ combine semantic profiles, compact codes, and continual
retrieval over a capability registry~\cite{guo2026agentdiscovery}.
\system connects task matching to exact-key responsibility and
posting-list retrieval, with a budget for network lookups.

\paragraph{Semantic retrieval over DHTs.}
pSearch organizes documents in a semantic overlay
~\cite{tang2003psearch}; LSH-on-DHT systems map semantic or
multidimensional neighborhoods to hash buckets
~\cite{zhu2007semanticdht,haghani2009distributedlsh}, and NearBucket-LSH
uses nearby buckets to improve recall per message
~\cite{kraus2015nearbucket}. LSH and multi-probe search supply the
underlying collision and probe mechanisms
~\cite{gionis1999lsh,lv2007multiprobe}. Coarse quantization with
residual encoding is established in centralized approximate retrieval
~\cite{jegou2011productquantization}. \system uses coarse cells and
residual codes to construct separate recall and precision keys, then
allocates a physical lookup budget across their ordered probes.

\paragraph{Distributed retrieval and structured resource discovery.}
PlanetP combines a gossip-replicated compact index with distributed
content search and ranking~\cite{cuenca2003planetp}. INS resolves
attribute--value descriptions, SWORD evaluates
multi-attribute resource constraints, and Mercury supports range
queries with overlay load balancing
~\cite{adjiewinoto1999ins,oppenheimer2004sword,bharambe2004mercury}.
\system encodes hard compatibility conditions in namespace labels
and matches task semantics within that scope. It retains Chord and
Kademlia's exact-key responsibility abstraction
~\cite{stoica2003chord,maymounkov2002kademlia}; posting-list services
provide multi-descriptor storage at the responsible peers.

\paragraph{Routing and storage defenses.}
Secure DHT work studies malicious routing peers
~\cite{castro2002secure}, while the Sybil attack shows why unconstrained
identities weaken open systems~\cite{douceur2002sybil}. Whanau addresses
Sybil-resilient DHT routing under a social-graph trust assumption
~\cite{lesniewski2010whanau}. These works protect routing and identity;
\system instead verifies whether a returned posting is authorized for
its logical key. DHT implementations and measurements examine latency,
throughput, replication, and Internet-scale routing behavior
~\cite{dabek2004dht,wang2013mainline}. OpenDHT provides
expiring values and storage-allocation controls
~\cite{rhea2005opendht}, while Beehive uses proactive replication
for skewed lookup demand~\cite{ramasubramanian2004beehive}.
Load-balancing schemes redistribute key ranges or items across peers
~\cite{karger2006loadbalancing}.
These mechanisms complement \system's certified publication, which
verifies descriptor-to-key consistency. Physical sharding redistributes
the resulting posting lists while preserving their logical keys and
charging each shard lookup to the query budget.
\endgroup

\section{\nsrevision{Discussion and Conclusion}}
\label{sec:discussion}
\begingroup\color{black}

\subsection{\nsclarify{Limitations and Future Work}}
\label{sec:discussion:nongoals}
\label{sec:discussion:limitations}
\label{sec:eval:threats}
\system targets requests for which suitable providers
must be identified beyond a requester's preconfigured service set.
Our experiments measure descriptor retrieval and the network execution
of discovery. They do not establish how frequently such requests arise
in deployed agent applications, or whether dynamically selected
providers improve task outcomes after integration and authorization
costs. The model admits both agent endpoints and service APIs, but
the measured workload uses API descriptions. Interface adapters and
application-level studies are needed to evaluate the broader setting.

\nsclarify{Exact-cosine recall measures fidelity to the chosen encoder. ToolBench
provides LLM-generated relevance judgments, and the augmented corpus
controls capability density; these results do not establish human
semantic correctness or natural capability frequencies. Namespace
policy labels follow deterministic rules and lack independent human
validation. Two effects remain unmeasured: embedding similarity
within versus across policy classes at fixed input/output contracts,
and how increasing $J$ changes the near--far pair hit-probability gap
beyond the residual-code fit in Section~\ref{sec:eval:theory}.}

The namespace registry and resolver define a deployment's compatibility
contracts. An unresolved alias can exclude a relevant provider before
semantic matching. Registry evolution and resolver coverage across
independently governed deployments remain to be evaluated. Encoder
and registry changes require coordinated configuration adoption
and descriptor republication because $\nu$ is part of each semantic key.

\nsclarify{The prototype implements certified posting storage and the live
lookup path measured in Section~\ref{sec:eval:network}. Protocol
simulations supply committee state serialization and controlled
routing or entry-path observations; they do not evaluate distributed
committee agreement, committee capture, or identity admission.
The analytic routing backend in Appendix~\ref{sec:scaling:cost} is
checked against held-out Kademlia simulation runs. Committee state
serialization, distributed descriptor updates, and authenticated
shard-manifest synchronization remain to be integrated and evaluated
for candidate availability under churn, with query and maintenance
costs. Section~\ref{sec:impl} specifies the measured execution path,
and Section~\ref{sec:poison:assumptions} states the certification
assumptions.}

\Needspace{6\baselineskip}
\subsection{Conclusion}
\label{sec:conclusion}
\nsclarify{We presented \system, a certified semantic index
for discovering agent-accessible capabilities from natural-language
task intent in P2P networks. Providers may expose these capabilities
through agent endpoints or service APIs.
Its key construction and probe schedule bound
per-descriptor publication and budget network lookups over exact-key
DHTs, while membership certificates make descriptor-to-key consistency
verifiable at storage and retrieval. On API descriptors, comparable
recall and candidate exposure can be achieved with fewer publication
keys and DHT lookups than tuned LSH\@. Our Go/libp2p prototype on cloud
virtual machines shows that these lookup reductions translate into
faster query completion on same-region and cross-region Internet
deployments. Controlled protocol experiments validate publication
enforcement and characterize availability and repair costs during
descriptor updates. \system thus connects semantic retrieval quality
to enforceable publication rules and explicit network budgets within
a decentralized discovery protocol.}
\endgroup

\label{arxiv:main-end}

\bibliographystyle{plain}
\bibliography{refs}

\clearpage
\appendix
\section{\nsrevision{Protocol Details}}
\label{app:protocol-details}
\begingroup\color{black}

\subsection{Complete Descriptor and Canonicalization}
\label{app:descriptor-format}

We use $d$ to denote a particular version of a capability descriptor
and $\bar d$ to denote the complete descriptor for that version. We
write the complete descriptor as the tuple
\begin{equation}
	\bar d = (\field{id}_d,\ \field{pk}_{p_d},\ \tau_d,\ \sigma_d,\ \psi_d,\ x_d,
	\ \field{meta}_d,\ \field{ptr}_d),
	\label{eq:descriptor}
\end{equation}
where $p_d$ is the provider that owns and publishes $d$. A provider is
an authenticated principal identified by its public key
$\field{pk}_{p_d}$. The field $\field{id}_d$ is a provider-local
capability identifier that remains fixed across versions of the same
capability; \nsrevision{$\tau_d$ is a protocol-level admission class,
with a common \code{generic} value when no additional admission
distinction is required; $\sigma_d$ identifies a canonical
interface-compatibility family, which groups descriptors with compatible
input/output requirements (e.g., accepting a department
and date and returning appointment records); and $\psi_d$ is a coarse
execution-policy class, such as \code{no-network} or \code{web-enabled}.
The field $x_d$ expresses task intent and capability details for
semantic encoding: the clinic service's supported
specialties and query functions are described here, together with
supported languages, example I/O, and domain tags such as
\code{outpatient-appointments}.}
The field $\field{meta}_d$ carries
freshness and version hints (e.g., the declared software version, model
revision, and last-update time), while $\field{ptr}_d$ locates a
copy of the complete descriptor that can be fetched again for
verification (e.g., through a stable HTTPS URI).

A deterministic canonicalization function $\canon(\cdot)$ fixes field
order and strips noise. Let $H$ denote the protocol's cryptographic hash
function. The descriptor commitment is
\(C_d=H(\canon(\bar d))\), as defined in Equation~\eqref{eq:commitment}.
After fetching the complete descriptor through $\field{ptr}_d$, the
requester applies the same canonicalization and hash functions and
compares the result with $C_d$ to verify that the retrieved content
matches the referenced descriptor version.
Here, $C_d$ commits to one descriptor version; it is not a stable
capability identifier. To link versions of the same capability, we
compute a lineage handle as
\begin{equation}
\iota_d=H(\text{``LINEAGE''}\|\field{pk}_{p_d}\|\field{id}_d)
\label{eq:lineage-handle}
\end{equation}
where the fixed tag separates lineage handles from other hash
outputs. The lineage handle binds a provider public key and a
provider-local capability identifier across descriptor versions.
Each certified descriptor version has a lineage epoch $e_d$, assigned
by its anchor committee. The requester deduplicates postings by
descriptor commitment $C_d$ and resolves updates and revocations by
\nsclarify{lineage handle $\iota_d$, as detailed in Appendix~\ref{sec:freshness}.}

\subsection{Namespace Registry}
\label{app:namespace-registry}

\nsrevision{A namespace defines the compatibility scope within which
semantic discovery operates. Its label records the admission,
interface, and execution-policy conditions that a provider must
satisfy for a request. Within that scope, the semantic sketch matches
task intent to capability descriptions. Providers offering different
capabilities can therefore share a namespace when these conditions
are compatible. Adding a capability under an existing compatibility
contract requires a new descriptor, without adding a namespace label.

Each deployment publishes a versioned namespace registry
$\mathcal R_\nu$ of canonical namespace labels. The registry uses a
small set of admission and policy classes, together with registered
interface-compatibility families. These are stable protocol
identifiers with a public byte encoding. An interface-family
identifier refers to a defined input/output contract; compatible
schema variants use the canonical family specified by the registry.
Descriptor $d$ has namespace label}
\(n(d)=(\tau_d,\sigma_d,\psi_d)\), as defined in Equation~\eqref{eq:namespace}.
\nsrevision{A field belongs in the namespace label when a mismatch makes
execution incompatible or violates policy. Task specializations and
other preferences remain in the semantic text and affect ranking.
For example, the clinic's appointment-query API in
\S\ref{sec:overview:example} may share a namespace with appointment
APIs for other specialties that use the same registered request/response
contract, admission class, and execution-policy class. Their semantic
texts distinguish the advertised specialties and functions. Merely
accepting and returning JSON does not establish interface compatibility.
The clinic's appointment assistant agent uses a different interface family
unless it exposes the same contract, directly or through an adapter.}

A query may admit several namespace labels. A deployment-supplied
resolver maps the requester's explicit hard requirements to an
allowed namespace-label set $\mathcal N_q\subseteq\mathcal R_\nu$;
the original task wording remains in $x_q$. A descriptor is eligible
exactly when $n(d)\in\mathcal N_q$. \nsrevision{For example,
the personal assistant's query using
the generic admission class, the illustrative
\code{appointment-query-v1} interface family, and the
\code{web-enabled} policy resolves to the label in
\S\ref{sec:overview:example}, while its eye-clinic query intent remains
in $x_q$. A client supporting multiple registered contracts may include
their labels in $\mathcal N_q$; discovery does not synthesize an adapter.}
The namespace registry and
resolver must agree across providers and requesters: an unresolved
alias or incompatible registry version can exclude a relevant
descriptor.

Namespace eligibility is encoded in the keys. A \emph{publication
key} is a logical key under which the provider publishes a posting;
a \emph{probe key} is a logical key accessed by the requester.
Publication-key construction includes $n(d)$, and probe-key
construction includes only labels in $\mathcal N_q$. Different
namespace labels therefore yield different logical keys even when
\nsclarify{coarse-cell IDs and residual codes match,
as specified in Section~\ref{sec:index:keygen}.} The requester also checks
$n(d)\in\mathcal N_q$ after fetching the complete descriptor.

The anchor committee validates the registered descriptor fields and
recomputes the namespace label before certifying the publication-key
\nsclarify{set, as described in Section~\ref{sec:poison:anchor}.}
When $\mathcal N_q$ contains multiple
labels, the requester constructs a probe sequence for each label and
\nsclarify{interleaves these sequences under one physical lookup budget $L_q$,
following the schedule in Section~\ref{sec:index:probe}.}

\subsection{Deterministic Probe Ordering}
\label{app:probe-order}

The requester computes $z_q=\enc_\nu(x_q)$ and orders the coarse-cell
IDs by distance to obtain $Z_q^{\mathrm{ord}}$. Its primary coarse-cell
set $Z_q$ contains the first $\rho_q$ IDs. For each considered coarse
cell $c$, the requester computes $r_{q,c}=z_q-\mu_c$ and residual
codes $b_{q,c}^{(j)}$ using the same construction as the provider.
Substituting $n\in\mathcal N_q$, $c$, and $b_{q,c}^{(j)}$ into
Equations~\ref{eq:recall-key} and~\ref{eq:precision-key} gives the
recall probe keys and precision probe keys.

The requester prioritizes selective precision probe keys before
accessing the larger posting lists at recall probe keys.
\nsrevision{The logical probe sequence has three stages: $P_1(q)$ for
top-cell precision probe keys, $P_2(q)$ for neighboring-code and
secondary-cell precision probe keys, and $P_3(q)$ for recall probe keys.}
$P_1$ contains exact precision probe keys for the primary cells and orders
them by cell rank and then residual-code-family index. For a configured
Hamming radius $r_H$ ($r_H=0$ disables this step), $P_2$ recovers
projection-boundary cases by substituting into
Equation~\ref{eq:precision-key} codes at Hamming distances $1$ through
$r_H$ from $b_{q,c}^{(j)}$, ordered by distance, cell rank, family index,
and residual-code bytes;
it then adds exact precision probe keys for cells ranked
$\rho_q+1,\ldots,\rho_{\mathrm{ext}}$, where $\rho_{\mathrm{ext}}$ is the
largest coarse-cell rank admitted for secondary-cell probing. $P_3$ orders recall probe keys by cell
rank. Within each stage, the per-label probe-key sequences are
interleaved round-robin in lexicographic namespace-label order. These
rules also provide deterministic tie breaking. The resulting logical probe sequence is
\begingroup\color{black}
the concatenation in Equation~\eqref{eq:logical-probe-sequence},
\endgroup
where $\Vert$ denotes sequence concatenation and \textsc{Dedup} keeps
the first occurrence of each probe key. In the single-namespace,
single-view, unsharded case, every probe key costs one physical lookup,
so the budgeted probe-key set is
\begingroup\color{black}
the prefix set in Equation~\eqref{eq:budgeted-probe-key-set},
\endgroup
where $\pref_L(S)$ returns
the first $L$ elements of sequence $S$. \nsclarify{As detailed in
Appendices~\ref{sec:poison:multiview} and~\ref{sec:scaling:cost}, the
general scheduler expands each probe key across entry views and
physical shards and charges every resulting lookup to the same $L_q$.}
A descriptor can be retrieved through a shared logical key when
$K_{\mathrm{pub}}(d)\cap K_{\mathrm{qry}}(q;L_q)\ne\emptyset$.

\subsection{LSH Key Construction}
\label{app:lsh-keys}

LSH-on-DHT maps embedding similarity to collisions at exact keys
~\cite{zhu2007semanticdht,haghani2009distributedlsh,kraus2015nearbucket}.
Let $\nu$ identify a shared semantic-index configuration, including
the encoder and namespace encoding. The baseline encodes descriptor
text as $z_d=\enc_\nu(x_d)$ and applies $T$ independently seeded LSH
functions $h_1,\ldots,h_T$. Descriptor $d$ is published at the keys
\begingroup\color{black}
\begin{equation}
\kappa_i^{\mathrm{LSH}}(d)=
H(\text{``LSH''}\|\nu\|n(d)\|i\|h_i(z_d)),
\label{eq:lsh-publication-key}
\end{equation}
\endgroup
for $i=1,\ldots,T$, where $H$ is a cryptographic hash function. A
requester computes $z_q=\enc_\nu(x_q)$ and, for each allowed namespace
label, probes the corresponding hash codes. Multi-probe LSH also
accesses neighboring codes~\cite{lv2007multiprobe}.

Hash width determines the selectivity of each posting list. Increasing
the width reduces candidate exposure but also reduces near-neighbor
collisions. More tables or neighboring-code probes can recover recall,
at the cost of additional publication or lookup work. Our evaluation
sweeps all three parameters and compares LSH-on-DHT with \system at
matched recall. Section~\ref{sec:index} develops the two-layer semantic
sketch used by \system to allocate coverage and discrimination to
separate keys.

\subsection{Registration and Certified Records}
\label{app:certified-records}

For lineage $\iota_d$, the stable anchor key is
$\kappa_A(\iota_d)=H(\text{``A''}\|\iota_d)$.
The anchor committee assigned to this key, $\anchor{\iota_d}$,
certifies successive descriptor versions without changing its
assignment when descriptor content changes. The provider submits
\begin{equation}
\mathrm{RegisterAnchor}(d)=\langle\iota_d,\bar d,C_d,\nu,e_d,\field{lease}_d,\field{sig}_{p_d}^{\mathrm{reg}}\rangle,
\label{eq:anchor-registration}
\end{equation}
where $\field{lease}_d$ is the requested expiration time and
$\field{sig}_{p_d}^{\mathrm{reg}}$ signs all preceding fields.
The anchor committee verifies the signature using the provider public
key in $\bar d$, checks $C_d=H(\canon(\bar d))$, derives $\iota_d$
from $\field{pk}_{p_d}$ and $\field{id}_d$
(Eq.~\eqref{eq:lineage-handle}), and validates the requested transition
against its lineage state.

The anchor committee then recomputes $n(d)$, $z_d$, and
$K_{\mathrm{pub}}(d)$ under configuration $\nu$. It constructs a
Merkle tree over the sorted publication-key set and issues the
membership certificate
\begingroup
\thinmuskip=1mu \medmuskip=1mu \thickmuskip=2mu
\begin{equation}
\begin{aligned}
\Cert_d={}&\langle\iota_d,C_d,\field{pk}_{p_d},\nu,n(d),e_d,\field{lease}_d,\field{root}_d,\field{prev}_d,\field{mode}_d\rangle,\\
\field{root}_d={}&\merkle(\mathrm{sorted}(K_{\mathrm{pub}}(d))),
\end{aligned}
\label{eq:membership-certificate}
\end{equation}
\endgroup
with committee signature $\Sigma_d$ over $\Cert_d$.
The provider public key is copied from the verified complete
descriptor. Carrying this key in $\Cert_d$ lets a posting-list service
verify a posting without first fetching $\bar d$.
The predecessor commitment $\field{prev}_d$ and state
$\field{mode}_d\in\{\code{live},\code{tomb}\}$
support the transitions in \S\ref{sec:freshness}. Supersession is a
local judgment derived from an observed successor; it does not modify
a signed membership certificate. Initial registration
uses $e_d=0$, $\field{prev}_d=\bot$, and
$\field{mode}_d=\code{live}$.

For each target logical key $\kappa$, the provider creates a posting body
\begin{equation}
\begin{split}
m_{d,\kappa}=\langle&\kappa,C_d,\field{pk}_{p_d},n(d),
\field{ptr}_d,\iota_d,e_d,\field{lease}_d,\\
&H(\Cert_d),\field{sig}_{d,\kappa}^{\mathrm{post}}\rangle,
\end{split}
\label{eq:minimal-posting}
\end{equation}
where $\field{ptr}_d$ locates the complete descriptor and
$\field{sig}_{d,\kappa}^{\mathrm{post}}$ signs all preceding fields.
The posting sent to the posting-list service is
the envelope in Equation~\eqref{eq:certified-posting},
where $\pi_{d,\kappa}$ proves $\kappa$'s inclusion under
$\field{root}_d$. Ordinary DHT routing locates the responsible peers;
their posting-list services perform acceptance and storage.

\subsection{Posting Verification and Quotas}
\label{app:posting-verification}

The posting-list service first binds the posting body to its
membership certificate and target logical key.
$\mathrm{Bind}(m_{d,\kappa},\Cert_d,\kappa)$ holds when the body's
target key equals the received logical key; its commitment, provider
public key, namespace label, lineage handle, epoch, and lease equal
the corresponding certificate fields; and its certificate reference
equals $H(\Cert_d)$.
Acceptance requires five checks:
\begin{equation}
\begin{aligned}
\mathrm{Accept}(d,\kappa)=1\iff{}&
\mathrm{Bind}(m_{d,\kappa},\Cert_d,\kappa)\\
&\wedge\,\mathrm{VerifySig}_{\field{pk}_{p_d}}(m_{d,\kappa})\\
&\wedge\,\mathrm{VerifyThresh}(\Cert_d,\Sigma_d)\\
&\wedge\,\mathrm{VerifyMerkle}(\field{root}_d,\kappa,\pi_{d,\kappa})\\
&\wedge\,\freshaccept(d,t_{\mathrm{now}}).
\end{aligned}
\label{eq:posting-list-acceptance}
\end{equation}
The provider-signature check uses the key in $\Cert_d$.
The committee-signature check verifies the assigned anchor committee,
threshold, and supported configuration $\nu$.
The lineage-aware predicate $\freshaccept$, defined in
\S\ref{sec:freshness:resolution}, uses the service's current time and
observed certified state. This predicate admits ordinary live postings.
The posting-list service verifies and stores tombstones and Move hints
using the lineage and historical key-membership checks in
\S\ref{sec:freshness:transitions}--\ref{sec:freshness:migration}.

An unauthorized target key fails the Merkle check. Repeated
publication is handled separately: the posting-list service retains
one active posting per $(C_d,\kappa)$, replacing it when a valid
monotonic renewal arrives. Replaying the same posting therefore does
not add entries to that posting list.

Distinct descriptor versions and lineages require broader accounting.
An optional quota $Q_{\mathrm{ns}}$ limits the simultaneously active
certified descriptor states of one admitted provider within a
namespace. Because the count spans lineages and anchor committees, the
deployment's admission or quota-accounting layer supplies an
authenticated quota-slot authorization before a new state is
certified. A renewal of unchanged state reuses its slot.
This quota bounds per-provider publication volume; its effect depends
on the admission policy in \S\ref{sec:poison:assumptions}.

\subsection{Multi-View Lookup and Ranking}
\label{app:multi-view}
\subsubsection{Probe Schedule and Provenance}
\label{sec:poison:multiview}

Membership verification detects invalid returned postings. To observe
differences in which valid postings are returned, the requester uses
entry points $E_q=\{u_1,\ldots,u_{|E_q|}\}$ drawn from distinct routing
zones or bootstrap neighborhoods. A \emph{view} is an entry-point and
probe-stage pair $(u,P_i)$, \nsrevision{giving the configured view set
$\views{q}=E_q\times\{P_1,P_2,P_3\}$.}
A \emph{lookup action} $(u,\kappa)$ probes logical key $\kappa$ through
entry point $u$. For each selected key, the requester issues the action
through each entry point before advancing to the next key. Entry
points therefore probe the same keys. With uniform lookup costs, these
keys form a common prefix of $\mathcal P(q)$, preserving the stage
order in \S\ref{sec:index:probe}.
Every action consumes physical lookup budget. In the unsharded case,
$|E_q|$ entries can repeat at most
$\lfloor L_q/|E_q|\rfloor$ keys within budget $L_q$.
Appendix~\ref{sec:scaling:cost} accounts for manifests and physical
shards. Repetition across entry points trades semantic probe depth
for observations of the same keys through different paths.

For each returned posting, the requester constructs a
\emph{provenance record}
\begin{equation}
\Pi_{d,\kappa,u}=\langle(u,P_i),\kappa,\mathcal B_{d,\kappa,u},\field{ts},H(\Cert_d)\rangle,
\label{eq:lookup-provenance}
\end{equation}
using its issued action, local receipt time $\field{ts}$, and the
authenticated identities of responsible peers whose posting-list
services returned the posting. These identities form
$\mathcal B_{d,\kappa,u}$; a responder's unverified account of other
peers does not add identities to the set.
The provenance record is maintained by the requester and is separate
from the posting.
Routes from different entry points may converge at the same responsible
peer, whose identity is counted once when records are merged.

\subsubsection{Coverage Features and Ranking}
\label{app:coverage-ranking}

The requester applies Eq.~\eqref{eq:posting-list-acceptance} to every
returned posting using its own time and observed lineage state.
Only accepted postings contribute candidates or coverage observations.
A failed check rejects that posting; it does not by itself establish
provider misconduct. In particular, the Merkle inclusion proof is
outside the provider-signed posting body and can be replaced by a
routing peer, while a valid old posting can be replayed without the
provider's participation.

The requester merges accepted postings by $C_d$ and fetches each
complete descriptor once. It checks the commitment, recomputes the
namespace label and lineage handle, compares them and the provider
public key with $\Cert_d$, and requires $n(d)\in\mathcal N_q$.
Descriptor versions are then resolved by lineage using
\S\ref{sec:freshness:resolution}.

This resolution distinguishes lease renewal from conflicting state.
Define the certified state identity
\begin{equation}
\begin{split}
h_{\mathrm{state}}(\Cert_d)=H(&\iota_d\|C_d\|\field{pk}_{p_d}\|\nu\|n(d)
\|e_d\|\field{root}_d\\
&\|\field{prev}_d\|\field{mode}_d).
\end{split}
\label{eq:certified-state-identity}
\end{equation}
A monotonic lease renewal preserves this identity; the full
$H(\Cert_d)$ in each provenance record identifies the particular
membership certificate returned.
If authenticated certificates claim incompatible states at the same
lineage epoch, the requester consults the stable anchor key and
withholds that lineage while the conflict remains unresolved.
Repeated observations of one state do not resolve the conflict.

For an eligible descriptor version $d$, let
$\views{q}(d)\subseteq\views{q}$ contain the views that returned an
accepted posting, and let
$\mathcal B_q(d)=\bigcup_{\kappa,u}\mathcal B_{d,\kappa,u}$
contain its distinct authenticated responders. Define
\begin{equation}
\cov_q(d)=\frac{|\views{q}(d)|}{3|E_q|},
\qquad
\rep_q(d)=\min\left(1,\frac{|\mathcal B_q(d)|}{b_0}\right),
\label{eq:coverage-features}
\end{equation}
where $b_0>0$ is the configured responder count at which responder
coverage saturates. View coverage uses the configured view set:
a stage that receives no probe budget contributes no observation.
Coverage therefore depends on both semantic overlap and the executed
probe schedule. Entry-path comparisons use matched $(P_i,\kappa)$
actions across entry points; disjoint keys would confound path effects
with semantic selection.

Among verified, namespace-eligible, current descriptor versions, the
requester ranks by
\begin{equation}
\mathrm{score}(d\mid q)=\lambda_{\mathrm{sem}}\mathrm{sim}(z_q,z_d)+\lambda_{\mathrm{cov}}\cov_q(d)+\lambda_{\mathrm{rep}}\rep_q(d),
\label{eq:final-score}
\end{equation}
with nonnegative weights. Certificate consistency and freshness
determine eligibility before this ranking step. An optional
provider-diversity cap limits how many descriptor versions from one
provider enter $\cand{q}$.
Appendix~\ref{sec:eval:multiview} evaluates the coverage signal and the
recall--cost trade-off of repeated probes.
\endgroup

\section{\nsrevision{Precision-Key Hit Analysis}}
\label{app:hit-analysis}
\begingroup\color{black}
\subsection{Precision-Key Hit Probability}
\label{sec:index:theory}

The two-layer sketch yields a precision-key hit through two successive
events: the query and descriptor first select at least one common
primary cell, and their residual codes then agree within the configured
Hamming radius in at least one residual-code family. Equivalently, for
some $c\in Z_q\cap Z_d$ and family $j$, a precision probe key derived by
the query exactly equals one of the descriptor's precision publication
keys. We analyze this primary-cell precision path before the physical
budget $L_q$ truncates the logical probe sequence. The model excludes
schedule-dependent secondary-cell and $P_3$ recall paths.

For a namespace-eligible query--descriptor pair, let
$\delta(q,d)=1-\mathrm{sim}(z_q,z_d)$. In this subsection, conditioning
on $\delta$ abbreviates conditioning on $\delta(q,d)=\delta$.
\nsrevision{Define the coarse-overlap event $A=\{Z_q\cap Z_d\ne\emptyset\}$,}
and let $B_J$ denote the event that, on at least one shared primary cell,
at least one of the $J$ residual-code families yields a matching
precision key. A precision-key hit on the primary-cell path occurs
when $A\cap B_J$ occurs.

\nsrevision{The coarse-cell overlap probability is
$g_{\rho,\rho_q}(\delta)=\Pr[A\mid\delta]$.}
Conditioned on a realized pair satisfying $A$, let $p(q,d)$ denote the
probability, over one randomly seeded residual-code family, that at
least one shared cell yields a matching precision key. This is a
per-pair probability: it depends on the realized residual geometry of
$q$ and $d$, rather than only on their embedding distance $\delta$. If
the $J$ families are approximately independent after conditioning on
that geometry, then
\begingroup\color{black}
\begin{equation}
\Pr[B_J\mid q,d,A]\approx 1-\bigl(1-p(q,d)\bigr)^J.
\label{eq:conditional-family-hit}
\end{equation}
\endgroup

Averaging over residual geometries and applying the probability product
rule gives the precision-key hit-probability model
\begin{equation}
\begin{aligned}
P_{\mathrm{hit}}(\delta)
 &= \Pr[A\cap B_J\mid\delta] \\
 &= \Pr[A\mid\delta]\Pr[B_J\mid A,\delta] \\
 &\approx g_{\rho,\rho_q}(\delta)\cdot\mathbb{E}\!\left[
 1-\bigl(1-p(q,d)\bigr)^J \,\middle|\, A,\delta\right].
\end{aligned}
\label{eq:hit-probability}
\end{equation}
The first two lines apply the probability product rule to coarse-cell
overlap and residual-code agreement. Only the final
line is approximate, because it models the $J$ residual-code families
as conditionally independent. The expectation remains outside the
nonlinear union because, in general,
$\mathbb E[1-(1-p)^J]\ne 1-(1-\mathbb E[p])^J$.

The single-family probability has a closed-form cell-level component.
For a shared cell $c$, let
$\theta_c(q,d)=\angle(r_{q,c},r_{d,c})$ be the angle between the query
and descriptor residuals. Under sign projection, one bit agrees with
probability $p_{\mathrm{bit}}(\theta)=1-\theta/\pi$. With $\ell$
independent projection bits, the probability that the two residual
codes differ in at most $r_H$ positions is
\begingroup\color{black}
\begin{equation}
p_{\ell,r_H}(\theta)=
\sum_{i=0}^{r_H}\binom{\ell}{i}
\bigl(1-p_{\mathrm{bit}}(\theta)\bigr)^i
p_{\mathrm{bit}}(\theta)^{\ell-i}.
\label{eq:residual-code-hit}
\end{equation}
\endgroup
This is exactly the probability that the descriptor's code lies within
the query's Hamming-$r_H$ precision probes for that cell; $r_H=0$
reduces it to the exact-code probability
$p_{\mathrm{bit}}(\theta)^\ell$. The pair-level $p(q,d)$ combines these
cell-level opportunities over all cells in $Z_q\cap Z_d$.

Increasing $J$ gives a near-neighbor pair more opportunities to
match a precision probe key, but also increases publication fan-out
and can expose more distant descriptors. The query budget further
limits how many of these opportunities are used.
Section~\ref{sec:eval:theory} evaluates the residual-code term and
the conditional-independence approximation on real residual pairs.
\endgroup

\section{Descriptor Updates and Freshness}
\label{sec:freshness}

A provider can renew a descriptor, publish a new descriptor version,
or revoke a capability. Each action changes which postings a requester
may use. This section defines these transitions, preserves
discoverability when publication keys change, and explains how
requesters resolve and propagate the resulting lineage state.

\Needspace{10\baselineskip}
\subsection{Lineage State}
\label{sec:freshness:lineage}
\label{sec:freshness:versioned}

The clinic service provider in
\S\ref{sec:overview:example} may revise its supported specialties
or interface contract while retaining the same capability identifier.
Changing appointment slots does not update the descriptor.
Let $d$ be its current descriptor version and $d'$ its successor.
The versions share $\iota_{d'}=\iota_d$, while the anchor committee
assigns $e_{d'}=e_d+1$ and sets $\field{prev}_{d'}=C_d$ in
$\Cert_{d'}$, using the membership certificate defined in
\S\ref{sec:poison:anchor}.

A membership certificate records the state signed by the anchor
committee. Its $\field{mode}$ is \code{live} for a descriptor version
or \code{tomb} for revocation. Observing a certified successor makes
an older live version \emph{superseded} in local lineage state;
neither the older membership certificate nor its signature changes.
The anchor committee serializes transitions under
\S\ref{sec:poison:assumptions}. Responsible peers at the stable anchor
key $\kappa_A(\iota_d)$ retain certified lineage state as publication
keys change.

Expiration times use a common time basis. The deployment bounds
clock error and includes that bound in its expiration checks; the
notation below treats those checks as comparisons against local
time. The anchor committee retains the highest epoch and any
revocation across lease expiry and committee changes. Posting-list
services likewise retain the highest epoch they have learned when
they remove expired postings.

\subsection{Renewal, Update, and Revocation}
\label{sec:freshness:transitions}

Table~\ref{tab:lineage-transitions} summarizes the three transitions.
In each case, the provider signs a request identifying the lineage,
the preceding membership certificate, and the requested change.

\begin{table}[t]
\centering\small
\begin{tabular}{@{}p{0.27\linewidth}p{0.68\linewidth}@{}}
\toprule
Transition & Certified change and publication action \\
\midrule
Lease renewal & Keep the epoch and certified state identity;
extend the lease and replace postings with newly signed posting bodies. \\
\addlinespace
Version update & Advance the epoch by one; certify the successor
and publish at its publication keys. \\
\addlinespace
Revocation & Advance the epoch by one; issue a tombstone and
disseminate it through the stable anchor and prior publication keys. \\
\bottomrule
\end{tabular}
\caption{Lineage transitions. Supersession is derived locally from
an observed successor.}
\label{tab:lineage-transitions}
\end{table}

\paragraph{Lease renewal.}
The anchor committee renews only its current live epoch. At approval
time $t_{\mathrm{approve}}$, it requires
\begingroup\color{black}
\begin{equation}
\begin{aligned}
t_{\mathrm{approve}}&<\field{lease}_{\mathrm{old}},\\
\field{lease}_{\mathrm{old}}&<
\field{lease}_{\mathrm{new}}\leq t_{\mathrm{approve}}+\Delta_{\max},
\end{aligned}
\label{eq:lease-renewal-conditions}
\end{equation}
\endgroup
where $\Delta_{\max}$ is the maximum lease duration. All fields in
$h_{\mathrm{state}}(\Cert_d)$
(Eq.~\eqref{eq:certified-state-identity}) remain unchanged.
The provider obtains the renewed membership certificate and committee
signature, then signs replacement posting bodies containing the new
lease and $H(\Cert_d)$. The publication-key set and Merkle inclusion
proofs remain valid, so renewal does not recompute the semantic sketch.

The pre-expiry condition applies when the anchor committee approves
the renewal. A replica receiving the renewed membership certificate
later may install it if the new lease is still valid and its local
state contains no higher epoch, revocation, or conflict. Expiry before
approval ends the renewal path; an admitted provider can request a
new epoch chained to the last certified state, with full
certification.

\paragraph{Version update.}
A change to the complete descriptor requires a new epoch. The anchor
committee checks
$\iota_{d'}=\iota_d$, $e_{d'}=e_d+1$, and
$\field{prev}_{d'}=C_d$, then recomputes the successor's namespace
label, semantic sketch, and publication-key set. The provider publishes
the resulting postings at $K_{\mathrm{pub}}(d')$. Participants that
observe $\Cert_{d'}$ mark lower live epochs superseded. If the
publication-key set changes, the provider also installs the Move hints
defined in \S\ref{sec:freshness:migration}.

\paragraph{Revocation.}
After authenticating the provider's request, the anchor committee
issues a tombstone using the membership-certificate schema. It retains
the preceding certificate's descriptor metadata, advances the epoch
by one, sets the predecessor commitment to $C_d$, and changes
$\field{mode}$ to \code{tomb}. The tombstone terminates that lineage:
the anchor committee accepts no later renewal or successor. Publishing
the capability again requires admission of a new lineage handle.
The lease field does not expire the revocation.

The provider publishes the tombstone and committee signature at
$\kappa_A(\iota_d)$. At a prior publication key $\kappa$, it additionally
supplies the prior membership certificate, committee signature, and
Merkle inclusion proof for $\kappa$. The posting-list service verifies
both certificates, the common lineage and provider key, the forward
epoch relation, and the old-key membership proof before storing the
tombstone and updating local lineage state. These checks authenticate
revocation independently of the live-posting acceptance predicate.
An expired prior lease does not invalidate its historical key-membership
evidence.

\subsection{Migration with Move Hints}
\label{sec:freshness:migration}

When an update changes the publication-key set, queries that still
probe the old keys need a way to discover the successor.
During a dual-publication window $\Delta_{\mathrm{mig}}$, the provider
publishes $d'$ at its publication keys and leaves a \emph{Move hint}
at each old publication key:
\begingroup\color{black}
\begin{equation}
\mathrm{Move}_{d\to d'}=\langle\iota_d,C_d,C_{d'},e_{d'},t_{\mathrm{end}},\field{sig}_{d\to d'}^{\mathrm{move}},\Cert_{d'},\Sigma_{d'}\rangle.
\label{eq:move-hint}
\end{equation}
\endgroup
Here $t_{\mathrm{end}}=t_{\mathrm{update}}+\Delta_{\mathrm{mig}}$
is an absolute expiration time, and the provider signature covers
all preceding fields. The provider chooses a window ending no later
than the successor certificate's lease. Including $\Cert_{d'}$ and
$\Sigma_{d'}$ makes the successor's certified state available to the
receiving posting-list service.

For storage at $\kappa\in K_{\mathrm{pub}}(d)$, the provider sends the
Move hint together with $\Cert_d$, $\Sigma_d$, and
$\pi_{d,\kappa}$. The posting-list service verifies the old-key
membership proof, both committee signatures, and the provider's Move
signature under the certified provider key. It also checks the shared
lineage, both commitments, consecutive epochs,
$\field{prev}_{d'}=C_d$, $\field{mode}_{d'}=\code{live}$, and
$t_{\mathrm{now}}<t_{\mathrm{end}}\leq\field{lease}_{d'}$.
The old certificate supplies historical membership evidence; the old
descriptor version need not remain live. The service stores the Move
hint through its expiration and returns it on probes of the old key.
The old descriptor version does not enter the candidate shortlist.

A requester verifies the Move hint and resolves the lineage through
the stable anchor key. It fetches the current complete descriptor,
checks its commitment and namespace eligibility, and admits the
descriptor only after the freshness checks below. Additional updates
are resolved by the same process. Anchor and successor-key lookups
consume the remaining physical lookup budget $L_q$; a resolution that
cannot finish within that budget yields no candidate for that lineage.
Appendix~\ref{sec:eval:migration} evaluates the window's effect on
discoverability.

\subsection{Freshness Resolution}
\label{sec:freshness:resolution}

Let $\Obs_q(\iota)$ be the membership certificates whose committee
signatures the requester has verified for lineage $\iota$, including
certificates learned through prior observations and anchor resolution.
The certificate accompanying an incoming posting is included after
authentication and before evaluating freshness. The set retains epoch
and revocation evidence after leases expire.
Writing $e(\Cert)$ for the epoch field of a membership certificate,
the highest observed epoch is
\begin{equation}
e_q^\star(\iota)=
\max\{e(\Cert):\Cert\in\Obs_q(\iota)\},
\label{eq:freshest-version}
\end{equation}
with $\max\emptyset=\bot$. This maximum is taken before checking
whether the highest epoch is live or its lease is valid.

Under the committee state-serialization assumption, a committee
signature authenticates a certificate's lineage and predecessor
relation. When combining certificates, the requester checks
predecessor links against observed state. A missing link between
observed epochs or conflicting certified state identities triggers a
stable-anchor lookup for the certificates needed to resolve that
lineage. Move hints, tombstones, and a deployment-defined near-expiry
threshold also trigger anchor resolution. Every such lookup is charged
to $L_q$.

Define $\Consistent_q(\iota)=1$ when the observed certificates have
compatible lineage and predecessor fields, no unresolved same-epoch
state conflict, and no successor after a tombstone. Compatible
same-epoch renewals have one certified state identity; local state
retains the largest verified lease. If required resolution leaves a
conflict or a predecessor gap unresolved, $\Consistent_q(\iota)=0$.
A repeated certificate does not outweigh conflicting certified state.

For a descriptor version $d$ with a verified membership certificate,
freshness at query time $t_q$ is
\begin{equation}
\begin{aligned}
\fresh_q(d)=1\iff{}&
\Consistent_q(\iota_d)=1\\
&\wedge\,e_d=e_q^\star(\iota_d)\\
&\wedge\,\field{mode}_d=\code{live}\\
&\wedge\,t_q<\field{lease}_d.
\end{aligned}
\label{eq:lineage-freshness}
\end{equation}
A posting-list service evaluates $\freshaccept(d,t)$ by the same rule
using its local observed certificates and time $t$. The lease check
uses the membership certificate bound to the presented posting body;
learning a longer renewal does not change an old body's signed fields.
Membership verification and descriptor reconstruction remain the
checks in \nsrevision{\S\ref{sec:poison:accept}}.

A highest epoch that is expired or a tombstone yields no live
candidate. An older version cannot become current again when that
highest epoch expires. Participants retain this epoch or revocation
information when deleting expired posting bodies. Freshness is
relative to observed certified state: a participant learns a new
update or revocation through publication, anchor resolution, or
read-repair.

\subsection{Read-Repair}
\label{sec:freshness:repair}

A requester that receives stale responses sends the newer membership
certificate and committee signature to the corresponding posting-list
services. It includes predecessor certificates needed to connect the
recipient's observed epoch to the newer state, or the recipient fetches
those certificates from the stable anchor. A certificate hash can
identify the update to fetch; installing the update requires the
certificate and its verifiable signature.

The receiving service verifies the committee signatures and predecessor
relations, advances its observed epoch, and marks older live versions
superseded. For a compatible renewal at the same epoch, it retains the
larger lease. Lower epochs do not replace higher epochs, and conflicts
follow the resolution rule in \S\ref{sec:freshness:resolution}.
An expired certificate can still establish that an older epoch has
been superseded; a tombstone continues to establish revocation.
A service admits a live posting only after the posting passes
Eq.~\eqref{eq:posting-list-acceptance}.

Repair messages disseminate certified state and do not reissue
membership certificates. DHT lookups performed by the requester for
repair consume $L_q$; direct repair messages and recipient-side
background fetches are maintenance traffic accounted for separately.
Appendix~\ref{sec:eval:freshness} reports stale exposure, convergence,
and repair-message cost.

\section{Scalability under Skew and Churn}
\label{sec:scaling}

Sections~\ref{sec:index}--\ref{sec:poison} and Appendix~\ref{sec:freshness}
define the logical index, certification, and freshness rules. Popular
semantic regions can nevertheless concentrate query, storage, and
maintenance load. This section separates those logical rules from
physical layout: sharding may redistribute the posting list of a
logical key, but it must not change the publication-key set or exceed
the query's physical lookup budget.

\subsection{Why Semantic Regions Hot-Spot}
\label{sec:scaling:hotspot}

A semantic index is more hotspot-prone than a classic exact-object
DHT for a structural reason: many unrelated queries and many unrelated
providers can legitimately converge on the same handful of coarse
cells simply because that region of the embedding space is popular,
which a content hash never causes. For any logical key $\kappa$ at
time $t$, let $\lambda_\kappa^{\mathrm{qry}}(t)$ and
$\lambda_\kappa^{\mathrm{mnt}}(t)$ be its query and maintenance
(renew/update/move/repair) arrival rates and $N_\kappa(t)$ its active
posting count. We do not need a physically exact cost model, only one
that lets the system tell apart \emph{why} a key is hot:
\begin{equation}
\Lambda_\kappa(t) = \alpha_q \lambda_\kappa^{\mathrm{qry}}(t)
+ \alpha_m \lambda_\kappa^{\mathrm{mnt}}(t) + \alpha_n N_\kappa(t).
\label{eq:bucket-load}
\end{equation}
Here $\alpha_q,\alpha_m,\alpha_n\geq0$ convert the three measured
loads into a common load score. This score
separates a posting-heavy logical key (long posting lists inflate bandwidth and
narrowing cost even at modest query rates) from a query-heavy logical key
(a popular semantic region hit repeatedly regardless of list length)
from a maintenance-heavy logical key (high provider churn or capability
drift saturates control traffic even when neither of the other two is
large). A key enters scaling mode once any of four thresholds trips,
\begin{equation}
\begin{aligned}
\mathrm{Hot}_\kappa(t) = 1 \iff\ & \Lambda_\kappa(t) > \Theta_{\mathrm{load}}
\ \vee\ N_\kappa(t) > \Theta_{\mathrm{size}} \\
& \vee\ \lambda_\kappa^{\mathrm{qry}}(t) > \Theta_{\mathrm{qry}} \ \vee\
\lambda_\kappa^{\mathrm{mnt}}(t) > \Theta_{\mathrm{mnt}},
\end{aligned}
\label{eq:hot-bucket}
\end{equation}
where $\Theta_{\mathrm{load}},\Theta_{\mathrm{size}},
\Theta_{\mathrm{qry}},\Theta_{\mathrm{mnt}}>0$ are deployment-configured
thresholds for the corresponding quantities. A key is therefore hot
when any one threshold is exceeded. Everything below is built to satisfy four constraints
simultaneously: change physical layout only, never logical keys or the
publication-key set;
never bypass the membership-certificate or freshness checks of
Section~\ref{sec:poison} and Appendix~\ref{sec:freshness}; trigger and recover locally
rather than rebalancing the whole overlay; and keep the physical lookup budget
$L_q$ an explicit, interpretable quantity even after sharding is
introduced.

\subsection{Physical Sharding without Semantic Rewriting}
\label{sec:scaling:shard}

The membership certificate of \S\ref{sec:poison:anchor} fixes
which publication keys a descriptor may occupy. Changing those keys
during sharding would invalidate the certified descriptor-to-key
relation. \system therefore
shards only the physical storage of a hot logical key $\kappa$ and never
changes the publication-key set. \nsrevision{The physical layout is
recorded in the shard manifest}
\begin{equation}
G_\kappa^{(\xi_\kappa)} = \big\langle \kappa,\ \xi_\kappa,\ s_\kappa,
\{\kappa^{[r]}\}_{r=1}^{s_\kappa},\ \field{rule}_\kappa,
\field{lease}_\kappa^{\mathrm{sh}},\ \Sigma_\kappa^{\mathrm{sh}}\big\rangle.
\label{eq:shard-manifest}
\end{equation}
\nsrevision{The manifest assigns $\kappa$ a layout epoch $\xi_\kappa$ and
$s_\kappa$ physical shard keys, derived as}
\begingroup\color{black}
\begin{equation}
\kappa^{[r]} = H(\text{``SHARD''} \| \kappa \| \xi_\kappa \| r).
\label{eq:physical-shard-key}
\end{equation}
\endgroup
Here $\field{rule}_\kappa$ is the deterministic
commitment-to-shard assignment rule instantiated below,
$\field{lease}_\kappa^{\mathrm{sh}}$ is the manifest-expiration time,
and $\Sigma_\kappa^{\mathrm{sh}}$ is the signature or quorum certificate
required by the deployment's shard-layout policy over all preceding
manifest fields. \nsrevision{For descriptor $d$, the service selects a shard
using the deterministic assignment rule}
\begingroup\color{black}
\begin{equation}
\operatorname{sid}_\kappa(C_d) = 1 + (H(C_d) \bmod s_\kappa).
\label{eq:commitment-shard-assignment}
\end{equation}
\endgroup
\nsrevision{It routes the posting to physical key
$\kappa^{[\operatorname{sid}_\kappa(C_d)]}$ without recomputing the
semantic sketch or requiring the provider to obtain a new membership certificate. The
corresponding acceptance predicate is}
\begingroup\color{black}
\begin{equation}
\begin{aligned}
\mathrm{ShardAccept}(d,\kappa^{[r]})=1\iff{}&
\mathrm{Accept}(d,\kappa)=1\\
&\wedge\,\mathrm{VerifyManifest}(G_\kappa^{(\xi_\kappa)})\\
&\wedge\,r=\operatorname{sid}_\kappa(C_d).
\end{aligned}
\label{eq:shard-acceptance}
\end{equation}
\endgroup
\nsrevision{This predicate is a strict conjunction: a posting must pass}
every check from \S\ref{sec:poison:accept} \emph{and} land in its
assigned physical shard, so sharding narrows how a valid posting is
stored without weakening what counts as valid. Here
$\mathrm{VerifyManifest}$ checks the manifest's signature or quorum
certificate, expiration time, logical key, and layout epoch. A hot key that stays
hot for a sustained split window $\Delta_{\mathrm{split}}$ is split into
\begingroup\color{black}
\begin{equation}
s_\kappa' = \min\!\left(s_{\max},\max\!\left(2,
\left\lceil\frac{\Lambda_\kappa(t)}{\Theta_{\mathrm{load}}}\right\rceil\right)\right)
\label{eq:shard-split-count}
\end{equation}
\endgroup
\nsrevision{shards, where}
$s_{\max}$ is the deployment shard cap. Merges use strictly lower
hysteresis thresholds held for a merge window $\Delta_{\mathrm{merge}}$
to avoid flapping between layouts.

\subsection{Two Caches That Are Never Authoritative}
\label{sec:scaling:cache}

Sharding by itself does not remove manifest traffic, so \system adds a
directory cache for shard layout and a small posting-window cache for
recently validated postings. Both are non-authoritative.
\nsrevision{The directory cache stores entries of the form}
\begingroup\color{black}
\begin{equation}
X_\kappa^{\mathrm{dir}}=\big\langle\kappa,\xi_\kappa,s_\kappa,\{\kappa^{[r]}\}_{r=1}^{s_\kappa},\field{ttl}_\kappa^{\mathrm{dir}},H(G_\kappa^{(\xi_\kappa)})\big\rangle.
\label{eq:directory-cache-entry}
\end{equation}
\endgroup
\nsrevision{Each entry carries the directory-cache}
expiration time $\field{ttl}_\kappa^{\mathrm{dir}}$ and the
shard-manifest digest $H(G_\kappa^{(\xi_\kappa)})$.
It is invalidated on expiration, a newer authenticated shard-manifest
digest learned through
anti-entropy or a shard response, or an epoch mismatch returned by an
attempted shard. A mismatch charges the attempted old shards, an
authoritative logical-key lookup, and the current shards. A posting-window
hit still re-runs membership-certificate and lineage checks before shortlist
admission: caching can shortcut finding a candidate, never trusting it.

This rule separates two guarantees that are easy to conflate. Even a
stale cache cannot make an invalid posting pass candidate verification.
It can, however, miss candidates until a newer shard-manifest digest is
observed or the TTL expires. Our simulation assumes that shard or
anti-entropy transport surfaces mismatches between authenticated
shard-manifest digests; it does not
measure that transport's delay. Without such a channel, TTL only bounds
layout staleness and the system cannot claim immediate knowledge of the
current shard manifest.

\subsection{Budget-Aware Probing after Sharding}
\label{sec:scaling:cost}

Once a logical key can be sharded, the assumption in
\S\ref{sec:index:probe} that one probe key costs one physical
lookup breaks: looking up a sharded logical key now means fetching a manifest
and then fanning out to some or all of its physical shards. \system
makes this explicit rather than silently absorbing it into $L_q$:
\begin{equation}
\mathrm{cost}_q(u,\kappa) =
\begin{cases}
1, & \kappa\ \text{unsharded}, \\
1+s_\kappa, & \text{sharded, directory cache miss}, \\
s_\kappa, & \text{sharded, directory cache hit},
\end{cases}
\label{eq:physical-probe-cost}
\end{equation}
The cost is charged separately for every entry point $u$; namespace-label
expansion has already produced distinct probe keys. The scheduler
admits a set of actions
$A_q(L_q)\subseteq E_q\times\mathrm{set}(\mathcal P(q))$
only if
\begin{equation}
\sum_{(u,\kappa)\in A_q(L_q)}\mathrm{cost}_q(u,\kappa)\le L_q.
\label{eq:total-physical-budget}
\end{equation}
The budgeted probe-key set is the projection
\begingroup\color{black}
\begin{equation}
K_{\mathrm{qry}}(q;L_q)=
\{\kappa:\exists u,\ (u,\kappa)\in A_q(L_q)\}.
\label{eq:physical-budget-key-projection}
\end{equation}
\endgroup
Thus $L_q$ remains the explicit bound promised in
\S\ref{sec:index:probe} after namespace-label expansion, path diversity, and
sharding. The scheduler preserves probe-stage precedence and repeats
each selected key across entry points as defined in
\S\ref{sec:poison:multiview}. It admits a key only when the combined
cost of its actions across $E_q$ fits the remaining budget.
When physical lookup costs differ, the scheduler can reprioritize
eligible keys within a stage using estimated marginal utility per
unit cost for each lookup action,
\begin{equation}
\begin{split}
\mathcal U_q(u,\kappa) = \frac{1}{\mathrm{cost}_q(u,\kappa)}\bigl(&
\alpha_{\mathrm{gain}}\,\widehat{\mathrm{gain}}_q(\kappa)
+ \alpha_{\mathrm{score}}\,\widehat{\Delta\mathrm{score}}_q(\kappa) \\
&- \alpha_{\mathrm{lat}}\,\widehat{\mathrm{lat}}_q(\kappa)
- \alpha_{\mathrm{poll}}\,\widehat{\mathrm{poll}}_q(\kappa)\bigr).
\end{split}
\label{eq:probe-utility}
\end{equation}
Here $\widehat{\mathrm{gain}}_q$ estimates new-candidate yield,
$\widehat{\Delta\mathrm{score}}_q$ shortlist-score improvement,
$\widehat{\mathrm{lat}}_q$ latency, and
$\widehat{\mathrm{poll}}_q$ historical invalid-candidate exposure; the
four estimates are normalized before weighting, and
$\alpha_{\mathrm{gain}},\alpha_{\mathrm{score}},
\alpha_{\mathrm{lat}},\alpha_{\mathrm{poll}}\geq0$
are their utility weights. The scheduler ranks keys by the
cost-weighted mean of $\mathcal U_q(u,\kappa)$ across $u\in E_q$,
which equals estimated total utility divided by total physical cost.
It completes the selected key's actions before selecting another key.
The requester stops after a completed group when $\mathrm{Stop}_q = 1$ iff $|\cand{q}| \ge k_q$ and the largest
score improvement among the most recent probe batch falls below
$\varepsilon_{\mathrm{stop}}$, the configured minimum meaningful
score improvement.

We validate the routing substrate this cost model sits on top of with
a 64-bit full-Kademlia implementation and a separately fit analytic
routing backend, blind-tested on $N_{\mathrm{net}}=10^5$ lookups never used for
fitting: across four churn profiles, the largest relative errors we
observe are $16.1\%$, $10.0\%$, and $12.8\%$ on p50/p95/p99 latency
respectively, and all 17 predeclared consistency gates pass. At
$L_q=32$ under a shard capacity selected to exercise all three cost
branches, a naive accounting that treats
$L_q$ as a semantic-probe count rather than a physical-lookup budget
lets $56.0\%$ of queries silently exceed their declared physical
budget; billing every probe through $\mathrm{cost}_q(u,\kappa)$ instead
keeps every single query at or under its declared budget, at a
measured cost of roughly $6.5$ fewer average lookups per query
relative to a cold-directory baseline once the directory cache is
warm.

\subsection{Load under Skew}
\label{sec:scaling:load}

We stress the sharding policy with a Zipf query distribution over the
4,096 most frequently probed real keys in our corpus, at skew
parameters from uniform up to $1.4$. At Zipf $1.2$, sharding drops the
single hottest node's absolute response-work from $17{,}555.9$ to
$6{,}713.9$, the max/mean response-work ratio from $191.6$ to $50.3$,
and the load Gini coefficient from $0.952$ to $0.906$, at a cost of
roughly $5.19$ additional physical lookups per logical query on
average. Sharding does not reduce every component of load: because a
query still has to visit every shard of a key it probes,
sharding does not divide a key's query arrival rate by the number of
shards---it primarily redistributes posting-list response work and
storage, and it strictly adds base lookup work rather than removing
it.

\subsection{Convergence under Churn}
\label{sec:scaling:churn}

Layout can go stale independently of descriptor freshness, so a
candidate is only admissible once both hold relative to the newest
signed state the requester has observed. Define
$\mathrm{FreshLayout}_q(\kappa)=1$ exactly when the requester has
verified the highest layout epoch for $\kappa$ among the shard manifests
it has observed:
\begin{equation}
\begin{aligned}
\mathrm{Admissible}_q(d,\kappa) = 1 \iff\ & \mathrm{FreshLayout}_q(\kappa)
= 1 \\
\wedge\ & \fresh_q(d) = 1,
\end{aligned}
\label{eq:cache-admissibility}
\end{equation}
\nsclarify{separating the newest certified descriptor state observed by
the requester, as defined in Appendix~\ref{sec:freshness}, from the
newest signed layout observed for its logical key.} Responsible
replicas exchange shard-manifest digest messages of the form
$\langle \kappa, \xi_\kappa,
H(G_\kappa^{(\xi_\kappa)}), \field{ts}_\kappa \rangle$, where
$\field{ts}_\kappa$ is the message timestamp, and repair on mismatch
or epoch lag using the same authenticated shard-manifest mechanism as ordinary
layout changes, and a newly responsible node bootstraps state---not
semantic truth, which remains defined entirely by
Section~\ref{sec:poison} and Appendix~\ref{sec:freshness}---from any online replica. In
a controlled evaluation with 128 real hot keys and 20 replicas, this
read-repair mechanism lowers the stale-layout exposure AUC from
$3.562$ to $1.448$ and the p95 convergence tail from $20.74$ to
$18.82$ time units at a representative turnover rate. Comparing an
unsafe cache design that trusts any cached posting against the safe
design of \S\ref{sec:scaling:cache} makes candidate-acceptance safety
concrete. Given the simulation's signal that two authenticated
shard-manifest digests mismatch, safe caching
accepts no injected invalid candidate (measured correctness $1.000$,
identical to no caching) against $0.5697$ for the unsafe variant, at a
$2.1\%$ fallback rate to the authoritative logical key $\kappa$. This result does
not measure the real transport latency of learning the newer
shard-manifest digest or
claim that a stale layout cannot temporarily reduce recall.

\section{Supplementary Evaluation and Reproducibility}
\label{app:supplementary-evaluation}

\subsection{\nsrevision{Network Replay: Inputs and Complete Conditions}}
\label{app:tierc-replay}
\begingroup\color{black}
\begin{table*}[t]
\color{black}
\centering\footnotesize
\setlength{\tabcolsep}{4pt}
\begin{tabular}{@{}llllrrr@{}}
\toprule
Placement & System & Schedule & Cache & Mean [95\% CI] (s) & P95 [95\% CI] (s) & RPC errors \\
\midrule
Same & \system & Parallel & Cold & 3.819 [3.490, 4.156] & 10.289 [8.949, 12.742] & 1 \\
Same & \system & Parallel & Warm & 3.009 [2.725, 3.315] & 8.879 [7.339, 10.099] & 0 \\
Same & \system & Barrier & Cold & 3.977 [3.618, 4.334] & 11.407 [9.195, 13.323] & 0 \\
Same & \system & Barrier & Warm & 2.918 [2.634, 3.215] & 8.751 [7.630, 9.779] & 0 \\
Same & LSH & Parallel & Cold & 13.897 [13.628, 14.157] & 19.721 [18.591, 20.157] & 0 \\
Same & LSH & Parallel & Warm & 13.565 [13.301, 13.843] & 18.929 [18.141, 19.786] & 0 \\
Same & LSH & Barrier & Cold & 13.829 [13.532, 14.126] & 19.735 [18.096, 20.764] & 1 \\
Same & LSH & Barrier & Warm & 13.565 [13.288, 13.839] & 18.865 [18.151, 19.943] & 4 \\
Cross & \system & Parallel & Cold & 7.038 [6.673, 7.419] & 14.178 [11.955, 16.882] & 0 \\
Cross & \system & Parallel & Warm & 6.836 [6.446, 7.230] & 15.093 [13.004, 17.329] & 2 \\
Cross & \system & Barrier & Cold & 9.717 [9.297, 10.168] & 18.311 [15.564, 20.135] & 0 \\
Cross & \system & Barrier & Warm & 9.094 [8.704, 9.524] & 15.946 [14.551, 19.378] & 0 \\
Cross & LSH & Parallel & Cold & 28.938 [28.644, 29.221] & 33.318 [32.588, 34.265] & 0 \\
Cross & LSH & Parallel & Warm & 29.044 [28.770, 29.312] & 33.734 [32.462, 34.263] & 0 \\
Cross & LSH & Barrier & Cold & 32.162 [31.863, 32.452] & 36.290 [35.610, 37.314] & 1 \\
Cross & LSH & Barrier & Warm & 32.130 [31.840, 32.419] & 36.569 [35.967, 37.276] & 0 \\
\bottomrule
\end{tabular}

\caption{All network-replay conditions for the archived 299-query
sample. Estimates retain frozen sample weights and all error-bearing
queries. Each interval resamples complete query clusters; errors count
application RPC failures, not failed logical lookups.}
\label{tab:tierc-conditions}
\end{table*}

\paragraph{Frozen inputs.}
The 500-query input bundle is sampled from the 8{,}087-query workload
before network execution. The \system configuration uses 16 coarse
centroids, $\rho=2$, $J=4$, $\ell=3$, $r_H=0$, and a probe-key budget
of 64. The LSH configuration uses 16 tables with eight-bit signatures,
Hamming radius two, and a probe-key budget of 256. Publication sets
and probe sequences are materialized from these configurations;
the replay does not interpolate between operating points. The
materializer freezes the rebuilt coarse codebook used for this input
bundle. Over the source workload, rebuilding changes eight candidate
incidences, while the median and P95 candidate counts and mean probe
count match the original configuration's saved results.

For each overlay, the materialized workload publishes 171{,}339
\system postings at 8{,}570 keys and 226{,}857 LSH postings at
42{,}749 keys. Setup readback checks 10{,}970 and 92{,}575 keys,
respectively, including planned keys with empty posting lists.
This check compares the returned descriptor commitments with the
materialized input before the timed replay begins.

\paragraph{Runtime settings.}
The run uses 200 peers per overlay, replication three, read and write
quorums two, page size 64, lookup concurrency 32, and 60-second
lookup and RPC timeouts. Replica completion uses the wait-all policy.
Certificate caches have 65{,}536 entries, are isolated between
overlays, and are cleared before each condition; warm conditions
then execute the same query's certificate prelude. The timing
records separate this prelude from lookup completion. This experiment
contains one requester and one deployment for each placement.

\paragraph{Measured sample and intervals.}
The September 15 archive includes sequences 0--4{,}783: 299 distinct
queries, each with one observation in every condition. We retain the
positive weights assigned in the frozen sample and normalize them
within this prefix. The effective sample size is 298.99. For each
bootstrap replicate, we sample 299 query IDs uniformly with replacement
and keep each sampled query's original weight and complete set of
conditions. We use 2{,}000 replicates, seed 20260910, and percentile
95\% intervals. Weighted P95 is the left inverse of the weighted
empirical CDF\@. Paired contrasts use the same sampled query IDs for
both systems or schedules. For the scheduling contrasts in
Table~\ref{tab:tierc-scheduling}, each replicate subtracts the parallel
estimate from the stage-barrier estimate. The recorded stage indices
and keys agree across all conditions for each query and system:
the schedules execute the same budgeted key sequence and reconstruct
the expected candidate set. These intervals are exploratory and have no
multiple-comparison adjustment or interim stopping rule.

Table~\ref{tab:tierc-conditions} reports all 16 conditions. A seeded
query permutation and a 16-condition balanced Latin square interleave
the executions. The 299-query prefix does not end at a complete
16-query block, so it retains some condition-order imbalance. The
prefix also need not reproduce every stratum's final allocation.
We preserve its original weights and include every complete query.
An earlier execution was interrupted after 37 complete queries;
the restarted execution repeats those query IDs. We retain the earlier
records separately and do not count them as additional independent queries.

\paragraph{Errors and snapshot integrity.}
The nine RPC errors consist of eight connection resets and one
stream-termination (GOAWAY) error, affecting seven query clusters.
There are no failed logical lookups, verification rejections, or
candidate-set mismatches. Excluding those seven complete query
clusters changes any condition's weighted mean completion time by
at most 0.111\,s. That calculation is a sensitivity diagnostic;
the main results retain the error-bearing queries.

All nine requester source-file hashes match the download manifest.
The checkpoint ends after query 299, and its detailed logs reconcile
to 690{,}256 logical lookups and 2{,}070{,}768 replica attempts.
The compressed files contain sealed members for these queries and
an unsealed tail from the next query. Analysis excludes this
uncheckpointed tail. The archive covers 299 of the 500 planned queries;
the original full-schedule and zero-RPC-error criteria were not met.

\paragraph{Resource instrumentation.}
During the latter part of replay, separate samplers read service-cgroup
CPU counters once per second on the requester and 16 data hosts.
Requester entry and exit probes additionally record CPU counters
around the timed lookup path, excluding certificate prewarming.
The markers align with 144 conditions from ten queries, including
eight complete 16-condition query blocks. The maximum discrepancy
between a marker window and its recorded lookup time is 2.61\,ms.
These CPU windows include service background work and cover a short
contiguous portion of the run. Instrumentation may add overhead;
the latency analysis retains both instrumented and uninstrumented
observations. The samplers did not collect resident memory.

\paragraph{Accounting.}
Application RPC counts include replica requests and continuation
pages; a logical lookup can generate several RPCs. Message bytes are
the requester's encoded request and response frames. Accumulated
validation time counts concurrent work and cannot be added to a
wall-clock time breakdown. The saved runtime CPU-capacity metric and
Go heap-allocation metric do not measure actual process CPU use and
RSS; neither is used for a resource-saving claim.
\endgroup

\subsection{\nsrevision{Validation Fixture and Record Sizes}}
\label{app:validation-fixture}
\begingroup\color{black}
The microbenchmark uses protocol v2, which binds the provider public
key in the membership certificate and admits an initial lineage epoch
of zero. This separate archived microbenchmark ran under Go 1.27.0 on Windows/amd64 with
16 logical processors. Measurements use CIRCL's portable BLS12-381
implementation, with public keys in G1, signatures in G2, and distinct
messages for the committee signers. Each reported operation is
repeated ten times with independent benchmark invocations.

Table~\ref{tab:tierc-sizes} gives canonical CBOR sizes for the
16-key, five-of-seven fixture. Each size measures the indicated
object independently; nested encodings need not equal the sum of
the standalone component encodings. Cache measurements reuse the
same verified certificate and committee signature across validations.
They do not measure batch verification of distinct certificates.

\begin{table}[!htbp]
\color{black}
\centering\small
\begin{tabular}{@{}lr@{}}
\toprule
Encoded object & Bytes \\
\midrule
Posting body & 253 \\
Provider signature & 64 \\
Membership certificate & 266 \\
Committee signature & 103 \\
Certificate with committee signature & 372 \\
Merkle inclusion proof & 143 \\
Complete posting & 839 \\
\bottomrule
\end{tabular}
\caption{Protocol-v2 fixture sizes. The complete posting includes
the signed body, certificate, committee signature, and inclusion proof.}
\label{tab:tierc-sizes}
\end{table}
\endgroup

\subsection{\nsrevision{Retrieval Configurations and Sensitivity}}
\label{app:retrieval-details}
\begingroup\color{black}
\paragraph{Codebook seeds and query uncertainty.}
Repeating the three representative points in
Table~\ref{tab:e2-operating-points} on two additional codebook seeds
keeps the qualitative ranking intact (recall stable within
$0.902$--$0.919$, $0.950$--$0.960$, and $0.970$--$0.978$
respectively), but exposure moves by 3--5 percentage
points across seeds, limiting the precision of a single-seed estimate.
For the primary seed, query bootstrap 95\%
confidence intervals for (recall, exposure) at the three
rows are respectively $(0.8987$--$0.9060,\ 0.6203$--$0.6311)$,
$(0.9471$--$0.9529,\ 0.7215$--$0.7335)$, and
$(0.9683$--$0.9727,\ 0.7957$--$0.8055)$.

\paragraph{Matched-recall selection.}
The operating point in Table~\ref{tab:e2-operating-points} at recall
target $0.95$ has publication fan-out eight. It is selected directly
by exposure among configurations already above that target.
Table~\ref{tab:baseline-headline} instead interpolates each
configuration's measured budget curve to recall exactly $0.95$
before minimizing exposure. The selected \system configuration then
has fan-out ten and slightly smaller interpolated exposure. The
network replay uses discrete configurations selected before execution,
as specified in Section~\ref{sec:eval:network}.

\begin{table}[t]
\centering\small
\begin{tabular}{@{}rrr@{}}
\toprule
& \multicolumn{1}{c}{\system} & \multicolumn{1}{c}{LSH-on-DHT} \\
recall target & exposure/probes & exposure/probes \\
\midrule
0.80 & 47.5\% / 30.6 & 47.1\% / 102.1 \\
0.90 & 62.1\% / 33.7 & 61.6\% / 173.2 \\
0.95 & 72.7\% / 32.9 & 72.7\% / 253.5 \\
0.97 & 79.0\% / 65.2 & 83.5\% / 110.2 \\
\bottomrule
\end{tabular}
\caption{\system versus the best-tuned LSH-on-DHT configuration at
matched recall, interpolated from measured budget curves. ``probes''
is the mean number of exact-key DHT lookups actually issued.}
\label{tab:baseline-headline}
\end{table}
\endgroup

\subsection{\nsrevision{Multi-View Lookup under Entry-Path Suppression}}
\label{sec:eval:multiview}
\begingroup\color{black}
We replay 8{,}087 queries with a one-entry budget of $L_q=128$.
For each query and entry point, the model independently captures the
entry path with probability $f\in\{0.05,\ldots,0.50\}$.
A captured path suppresses a coordinated target set comprising
approximately $80\%$ of reachable descriptors and returns the
remainder. Here $f$ is an entry-path capture probability; the model
does not map it to a fraction of malicious DHT nodes.

We compare two probe schedules. The first repeats the complete
one-entry probe prefix at every entry, increasing total physical
cost. The second divides the original query's actual probe count
across entries and repeats the resulting shorter prefix at each
entry. Table~\ref{tab:multiview-budget} reports both schedules at
$f=0.20$. Repeating the complete prefix through two entries raises
attacked recall from $0.7957$ to $0.9191$, with twice the lookup cost.
Keeping mean cost at $59.1$ lookups reduces clean recall from
$0.9500$ to $0.8553$, as fewer distinct keys are probed.

\begin{table}[t]
\centering\small
\begin{tabular}{@{}lrrr@{}}
\toprule
Schedule & Lookups & \multicolumn{2}{c}{Recall} \\
& & Clean & Attacked \\
\midrule
One entry & 59.1 & 0.9500 & 0.7957 \\
Two, full prefix & 118.3 & 0.9500 & 0.9191 \\
Two, fixed cost & 59.1 & 0.8553 & 0.8273 \\
\bottomrule
\end{tabular}
\caption{Mean physical lookups and recall at entry-path capture
probability $f=0.20$. Both two-entry schedules repeat the same keys
across entries; the fixed-cost schedule uses a shorter prefix.}
\label{tab:multiview-budget}
\end{table}

For each query, we average $\cov_q(d)$ and $\rep_q(d)$ over returned
candidates and use one minus the mean of these two averages as the
query-level anomaly score. View coverage uses the configured $3|E_q|$
denominator in Eq.~\eqref{eq:coverage-features}. The simulation assigns
two distinct responder observations to each successful key--entry
lookup, with saturation $b_0=20$; it does not model routes converging
on the same authenticated responder.

The ROC analysis distinguishes clean queries from queries with at
least one captured entry path. At $f=0.20$, two entries repeating the
complete prefix yield AUC $0.928$ from view coverage, $0.892$ from
responder coverage, and $0.920$ from the combined score, compared with
combined AUC $0.496$ for one entry. When all entry paths are captured,
targeted candidates disappear from every view and the combined AUC
returns to approximately $0.5$. The experiment measures query-level
detection and coverage; the candidate-ranking weights in
Eq.~\eqref{eq:final-score} remain uncalibrated.
\endgroup

\begingroup\color{black}
\begin{samepage}
\subsection{\nsrevision{Storage under Skew and Layout Changes}}
\label{app:storage-sweeps}
Sharding's benefit under query skew is not limited to the single Zipf
level reported in \S\ref{sec:scaling:load}: across the full swept
range from uniform traffic to Zipf $1.4$, the gap between sharded and
unsharded hottest-node response-work widens as skew increases, from
a $1.5\times$ reduction under uniform traffic to $3.8\times$ at Zipf
$1.2$ and $4.9\times$ at Zipf $1.4$, the highest skew we tested.
\end{samepage}

Figure~\ref{fig:e9} presents the full skew sweep.

\begin{figure}[t]
\centering
\includegraphics[width=0.92\linewidth]{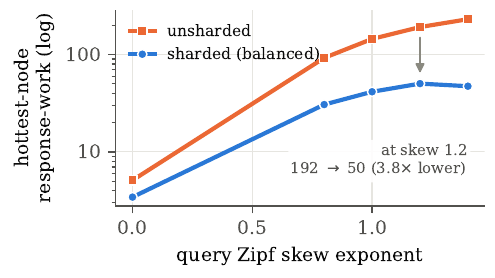}
\caption{Hottest-node response-work versus query Zipf skew, sharded
(balanced policy) versus unsharded (log-scale $y$-axis). The benefit
grows with skew; the arrow marks the 192-to-50 reduction at Zipf
$1.2$.}
\label{fig:e9}
\end{figure}

Meanwhile, caching with certificate re-validation keeps measured
candidate-acceptance correctness at $1.000$, versus $0.5697$ for a
\nsclarify{design that trusts cached postings without re-validation,
as reported in Appendix~\ref{sec:scaling:churn}.} The safe-cache result additionally assumes
that mismatches between authenticated shard-manifest digests reach readers; we do not
measure mismatch-transport delay or the temporary recall loss of a
reader that has not yet learned the new layout.
\endgroup

\subsection{Descriptor Migration}
\label{sec:eval:migration}
We construct 512 same-lineage, same-namespace reindexing cases from
the augmented corpus, selecting old and new descriptor versions with
disjoint publication-key sets. Both sets are reached by real query
probes at $L_q=128$. The resulting 20{,}276 query observations receive
controlled switch times from old to new keys within one normalized
rollout unit. We use uniform, front-loaded, and back-loaded switch
profiles. During an active migration window, a Move hint is modeled
as successful resolution of an old-key lookup; the metric measures
discoverability under this model and does not charge additional
anchor-resolution lookups.

\begin{figure}[t]
\centering
\includegraphics[width=0.85\linewidth]{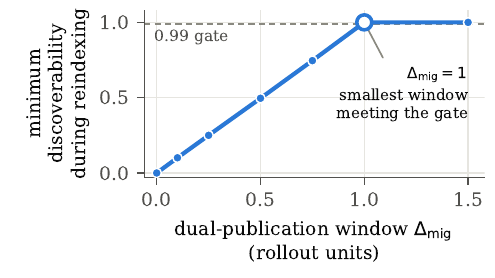}
\caption{Minimum discoverability in the controlled reindexing model
under uniform rollout. Old and new publication-key sets are disjoint;
query observations switch within one rollout unit. A one-unit Move
window covers that interval.}
\label{fig:migration}
\end{figure}

\nsrevision{We sweep the dual-publication window $\Delta_{\mathrm{mig}}$
over $\{0,0.1,0.25,0.5,0.75,1,1.5\}$.}
Without Move hints, minimum discoverability is $0$ at update time:
the selected key sets are disjoint and the modeled observations have
not yet switched. A window covering the full rollout reaches minimum
discoverability $1.000$ under all three profiles. This follows the
model's bounded switch times and successful Move resolution.
Shorter windows preserve discoverability initially and reduce the
subsequent loss. Under uniform rollout, a half-unit window raises
minimum discoverability to $0.496$ and lowers the integrated
discoverability shortfall from $0.502$ to $0.129$.
Figure~\ref{fig:migration} shows the window sweep.

The one-unit window places 4{,}096 Move hints at old publication keys,
with total residence time of 4{,}096 hint--rollout units.
Extending the window to $1.5$ raises residence time to $6{,}144$
without improving discoverability in this workload.

\subsection{Freshness and Read-Repair}
\label{sec:eval:freshness}
For silent provider loss, we use 512 lineages and 14{,}019 query-weighted
observations with controlled residual lease phases. Stale exposure is
the weighted fraction of postings whose leases remain valid after the
provider disappears. With a one-unit lease, the stale-exposure integral
is $0.499$ and the p95 stale window is $0.951$ units; exposure reaches
zero by one unit. Plain TTL with the same timeout and phases gives the
same curve. In a separate replay experiment, lineage checks reject
older postings after a verified update or tombstone; plain TTL
retains them until expiration.

The read-repair experiment starts each of 512 lineages with two of
eight responsible replicas holding the new epoch. Each query contacts
three distinct replicas, with query rates weighted by corpus query
hotness. Both experiment arms retain background anti-entropy at
$0.25$ per stale replica per time unit. Responsibility turnover at
rates $0$, $0.05$, $0.10$, $0.20$, and $0.40$ per time unit replaces
a replica's state with a randomly selected peer's state.
We run three seeds over a 40-unit horizon with a 0.05-unit step,
using perfect routing.

Here stale exposure is the query-weighted fraction of responsible
replicas still holding the old epoch. It measures replica state before
candidate verification. Convergence time is the first time all eight
replicas of a lineage hold the new epoch. For both freshness experiments, the
stale-exposure integral is the area under the corresponding exposure
curve over the observation horizon and has units of time; the
suppression detector's ROC-AUC is a separate metric.
Table~\ref{tab:read-repair-cost} reports the representative turnover
rate $0.10$.

\begin{table}[t]
\centering\small
\begin{tabular}{@{}lrrr@{}}
\toprule
Read-repair & \shortstack{Stale-exposure\\integral} & p95 time & Messages \\
\midrule
Off & 3.037 & 17.96 & 0 \\
On & 1.288 & 8.95 & 1,538 \\
\bottomrule
\end{tabular}
\caption{Read-repair at turnover rate $0.10$. Times are normalized
units; messages count direct read-repair messages for the
512-lineage run, averaged over three seeds. Both arms include
background anti-entropy.}
\label{tab:read-repair-cost}
\end{table}

\begin{samepage}
Across all five turnover rates, read-repair lowers the stale-exposure
integral from $2.955$--$3.089$ to $1.246$--$1.317$ and the p95
convergence time from $16.04$--$19.26$ to $8.66$--$9.30$ units,
a $1.84$--$2.08\times$ reduction in the convergence tail.
The experiment models propagation of authenticated state; validating
the complete control-record transport and recovery paths remains
implementation work.

\end{samepage}

\begin{table*}[t]
\centering\small
\setlength{\tabcolsep}{4pt}
\begin{tabular}{@{}lllrrrr@{}}
\toprule
& & & \multicolumn{2}{c}{CPU (\%)} & \multicolumn{2}{c}{RSS (MiB)} \\
\cmidrule(lr){4-5}\cmidrule(l){6-7}
System & Scheduler & Cache & Baseline & Query phase [range] & Baseline & Query phase [range] \\
\midrule
\system & Parallel & Cold & 0.399 & 0.326 [0.323, 0.328] & 51.87 & 52.17 [52.09, 52.25] \\
\system & Parallel & Warm & 0.411 & 0.291 [0.288, 0.296] & 52.25 & 52.16 [51.76, 52.44] \\
\system & Stage-barrier & Cold & 0.414 & 0.336 [0.326, 0.354] & 52.12 & 52.50 [52.29, 52.62] \\
\system & Stage-barrier & Warm & 0.416 & 0.294 [0.284, 0.307] & 51.76 & 51.68 [51.47, 51.87] \\
\midrule
LSH & Parallel & Cold & 0.409 & 0.394 [0.382, 0.408] & 52.44 & 52.85 [52.58, 53.11] \\
LSH & Parallel & Warm & 0.401 & 0.440 [0.431, 0.449] & 52.55 & 53.00 [52.90, 53.19] \\
LSH & Stage-barrier & Cold & 0.407 & 0.410 [0.394, 0.433] & 52.24 & 52.68 [52.39, 52.93] \\
LSH & Stage-barrier & Warm & 0.398 & 0.431 [0.421, 0.449] & 52.37 & 52.89 [52.62, 53.10] \\
\bottomrule
\end{tabular}

\caption{Per-node resource observations in the separate 400-node
deployment. Baseline and query-phase values are means of three
rounds, each averaging 25 cloud nodes. Brackets give the minimum
and maximum of the three query-phase batch means.
CPU 100\% denotes one logical core.}
\label{tab:tierc-resources-details}
\end{table*}

\subsection{Node Resource Measurements}
\label{app:tierc-resources}

The resource experiment in Section~\ref{sec:eval:resources} measures
25 colocated cloud node processes in a 400-node overlay. The local
server has an Intel Core i9-10980XE with 36 logical CPUs and about
251\,GiB of OS-visible memory. The Hong Kong VM has four Intel Xeon
Platinum vCPUs and 8\,GB RAM\@. The requester and controller run on
the local server. The measured resource values are for the cloud
node processes. Publication materializes the selected 50-query
workload: 39{,}631 \system postings at 1{,}228 keys and 46{,}503 LSH
postings at 6{,}489 keys. Each system retains its frozen query plans
and the replica, quorum, paging, and lookup-concurrency settings
from Section~\ref{sec:eval:network}.

The eight conditions combine the two systems, parallel and
stage-barrier scheduling, and cold and warm requester certificate
caches. Each condition runs once in each of three rounds, with the
same 50 query IDs and query order. Condition order is randomized
within each round. These repetitions provide 24 complete batches.
An interrupted 16-query attempt is retained separately and excluded
from every resource summary; its replacement supplies the complete
batch. Cold and warm refer to the requester certificate cache;
operating-system file caches are not reset.

The sampler reads cgroup v2 CPU usage and main-process RSS from
\texttt{/proc} at nominal 10-second intervals. It tracks process
identity and phase transitions. We exclude intervals crossing phase
boundaries and use actual elapsed time for CPU rates and weighted
RSS means. The complete batches contain 57{,}325 node samples with
no recorded sampling errors, process-identity changes, or CPU-counter
resets. Recomputing CPU rates from the raw counters reproduces the
recorded rates. Table~\ref{tab:tierc-resources-details} reports
baseline means and variation among the three query-phase batch
means. \nsclarify{The resource summaries retain all 1{,}200 executions,
including 493 that returned fewer candidates than expected, together
with their waiting and recovery work.}
Time samples and colocated nodes are not independent
experimental repetitions. The query phase includes certificate
prewarming and gaps between queries, so it does not isolate
per-query CPU cost.

\end{document}